\documentclass[journal]{IEEEtran}
\usepackage{caption}
\usepackage[pdftex]{graphicx}
\usepackage{subfig}
\usepackage[ruled]{algorithm2e}
\usepackage{xcolor,soul,framed} 
\usepackage{cite}
\usepackage{graphicx}
\colorlet{shadecolor}{yellow}

\graphicspath{{../pdf/}{../jpeg/}}
\DeclareGraphicsExtensions{.pdf,.jpeg,.png}

\usepackage[cmex10]{amsmath}
\usepackage{array}
\usepackage{mdwmath}
\usepackage{mdwtab}
\usepackage{eqparbox}
\usepackage{url}
\usepackage{multirow}
\usepackage{amssymb} 
\usepackage{svg} 
\usepackage{comment}

\usepackage{booktabs} 

\usepackage[colorlinks=true, citecolor=black]{hyperref}
\DeclareCaptionFont{blue}{}

\begin{document}
\bstctlcite{IEEEexample:BSTcontrol}
    \title{RadioGAT: A Joint Model-based and Data-driven Framework for Multi-band Radiomap Reconstruction via Graph Attention Networks}
  \author{Xiaojie~Li, Songyang Zhang,~\IEEEmembership{Member,~IEEE}, Hang Li, Xiaoyang Li,~\IEEEmembership{Member,~IEEE}, Lexi Xu, Haigao Xu, Hui Mei,
 Guangxu Zhu,~\IEEEmembership{Member,~IEEE}, Nan Qi,~\IEEEmembership{Senior Member,~IEEE}, and Ming Xiao,~\IEEEmembership{Senior Member,~IEEE}
     
\thanks{Xiaojie Li is with the Shenzhen Research Institute of Big Data, The Chinese University of Hong Kong-Shenzhen, Guangdong, China, and is also with the College of Physics, Nanjing University of Aeronautics and Astronautics, Nanjing 210016, China (e-mail: xiaojieli@nuaa.edu.cn).

Songyang Zhang is with the Department of Electrical and Computer Engineering, University of Louisiana at Lafayette, Lafayette, LA, 70503 (e-mail: songyang.zhang@louisiana.edu).

Hang Li, Xiaoyang Li and Guangxu Zhu are with the Shenzhen Research Institute of Big Data, The Chinese University of Hong Kong-Shenzhen, Guangdong, China (e-mail: \{hangdavidli, lixiaoyang, gxzhu\}@sribd.cn).

Lexi Xu is with the Research Institute, China United Network Communications Corporation, Beijing 100048, China (e-mail: davidlexi@hotmail.com).

Haigao Xu and Hui Mei are with China Mobile Communications Group Jiangxi Co., Ltd 330038 (email: \{xuhaigao, meihui\}@jx.chinamobile.com).

Nan Qi is with the Key Laboratory of Dynamic Cognitive System of Electromagnetic Spectrum Space, Nanjing University of Aeronautics and Astronautics, Nanjing 210016, China, and also with the State Key Laboratory of Integrated Services Networks, Xidian University, Xi’an 710071, China (e-mail: nanqi.commun@gmail.com).

Ming Xiao is with the Department of Information Science and Engineering, School of Electrical Engineering of KTH, Royal Institute of Technology, Stockholm, Sweden (e-mail: mingx@kth.se).
}
}

\markboth{IEEE TRANSACTIONS ON WIRELESS COMMUNICATIONS
}{Roberg \MakeLowercase{\textit{et al.}}: 
RadioGAT: A Joint Model-based and Data-driven Framework for Multi-band Radiomap Reconstruction via Graph Attention Networks}

\maketitle

\begin{abstract}
Multi-band radiomap reconstruction (MB-RMR) is a key component in wireless communications for tasks such as spectrum management and network planning. However, traditional machine-learning-based MB-RMR methods, which rely heavily on simulated data or complete structured ground truth, face significant deployment challenges. These challenges stem from the differences between simulated and actual data, as well as the scarcity of real-world measurements. To address these challenges, our study presents RadioGAT, a novel framework based on Graph Attention Network (GAT) tailored for MB-RMR within a single area, eliminating the need for multi-region datasets. RadioGAT innovatively merges model-based spatial-spectral correlation encoding with data-driven radiomap generalization, thus minimizing the reliance on extensive data sources. The framework begins by transforming sparse multi-band data into a graph structure through an innovative encoding strategy that leverages radio propagation models to capture the spatial-spectral correlation inherent in the data. This graph-based representation not only simplifies data handling but also enables tailored label sampling during training, significantly enhancing the framework's adaptability for deployment. Subsequently, The GAT is employed to generalize the radiomap information across various frequency bands. Extensive experiments using raytracing datasets based on real-world environments have demonstrated RadioGAT's enhanced accuracy in supervised learning settings and its robustness in semi-supervised scenarios. These results underscore RadioGAT's effectiveness and practicality for MB-RMR in environments with limited data availability.
\end{abstract}


%
\IEEEpeerreviewmaketitle


\section{Introduction}
 \IEEEPARstart{N}{ext-generation} wireless networks, such as beyond fifth-generation (B5G) and sixth-generation (6G) standard cellular networks, are expected to support the massive connection by the proliferation of intelligent terminals, as well as the demand of diverse ultra-wideband services and high-accuracy sensing\cite{b1,b2,b3}. To this end, there are a series of design and control issues to be addressed, such as network planning\cite{b4}, resource allocation\cite{b5}, dynamic spectrum access\cite{b6}, and localization\cite{b7}. All require high-precision radiomaps to facilitate the accurate assessment of the radio environment\cite{b8,b9,b10,b11}. Radiomap, which describes the distribution of power spectral density (PSD) information over geometric locations, time and frequencies, is usually reconstructed from a set of sparse observations collected by deployed sensors and mobile devices. However, how to reconstruct the high-precision full-band radiomap from its sparse samples in space, time and frequency band represents a longstanding challenge yet to be overcome and thus is the main theme of this paper.

\begin{figure}[t]
\centering
\includegraphics[width=6cm]{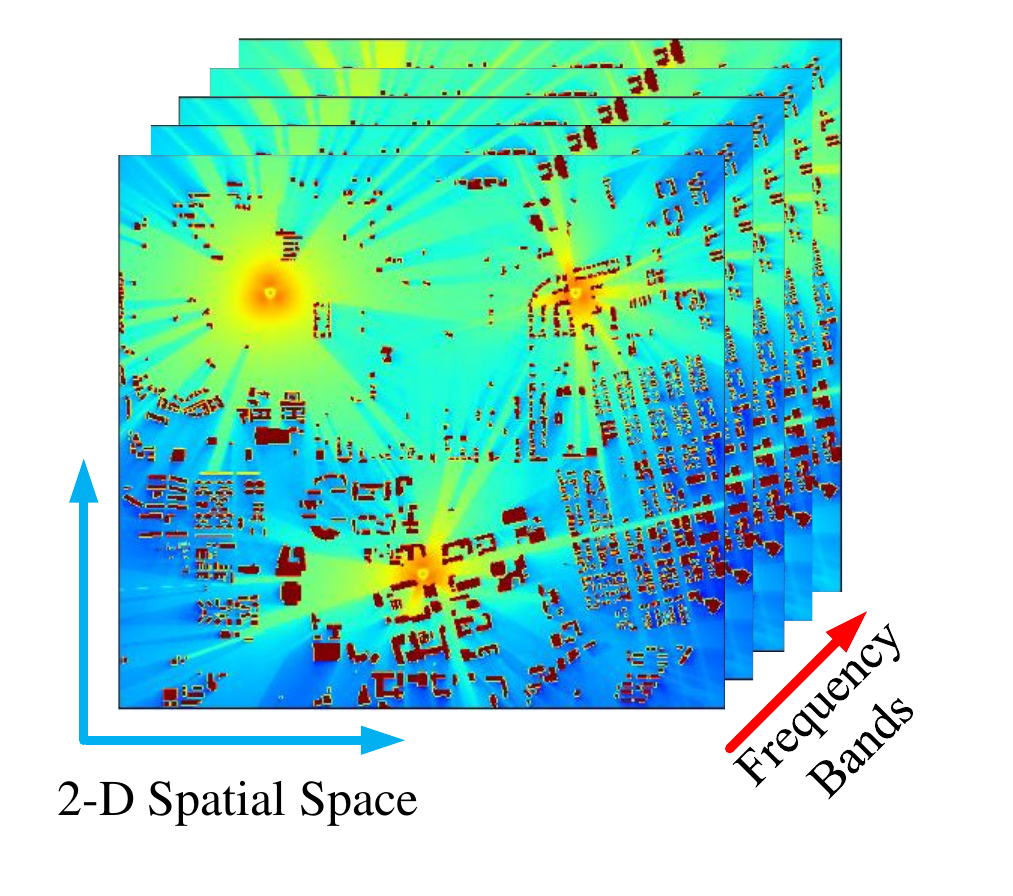}
\caption{The illustration of multi-band radiomap which includes three transmitters with dark red parts for buildings, and other parts for the strength of {received signal strength (RSS).}}
\label{fig1}
\end{figure}

Most of the traditional radiomap reconstruction  (RMR) methods predict PSD in the two-dimensional (2D) spatial spaces with a given frequency band (uni-band), focusing on the spectrum distribution over different locations. For example, one can apply radio propagation model to interpolate sparse radiomaps \cite{6150651} while also applying data-driven approaches to predict spatial radiomaps from surrounding observations \cite{b13}. More spatial radiomap approaches will be reviewed in Section \ref{related}. However, the recent demand of full band information in many modern wireless applications,
such as spectrum sharing, dynamic spectrum access, and ultra-wideband communications, triggers the development of constructing multi-band radiomaps over a wide range of spectrum of interests \cite{b31,b32}. As shown in Fig. \ref{fig1}, the multi-band radiomap represents the spatial-spectral distribution of spectrum occupancy over a regular map. Particularly, the colour of each position indicates the received signal strength (RSS) or the large-scale fading over different frequencies. Thus, one particular challenge lies on the characterization of spatial-spectral information in multi-band radiomap reconstruction (MB-RMR).

Existing works of RMR can be categorized into two groups: model-based and data-driven methods. In model-based approaches, missing values are usually estimated based on a fixed signal propagation model, with which the entire radiomap can be constructed \cite{b12}. However, due to the inevitable difference between the assumed model and a realistic scene, model-based methods often suffer from low construction accuracy under complicated surrounding environments \cite{10001269}. On the other hand, data-driven methods utilize data statistics to predict the missing values in radiomaps. With sufficient training data samples, the data-driven methods usually succeed in capturing the overall PSD distribution \cite{b13}. Many deep learning approaches, such as Feedforward Neural Networks \cite{b14}, Convolutional Neural Networks (CNN) \cite{b15}, and Generative Adversarial networks (GAN) \cite{b16}, have been applied for high-accuracy RMR. The efficacy of data-driven methods is intrinsically linked to the quality and quantity of the training datasets. In numerous instances, particularly within geographically constrained environments like mountainous terrains and regions afflicted by disasters, acquiring an adequate volume and quality of training samples is often an insurmountable challenge. To circumvent this limitation, researchers have turned to synthetic datasets sourced from raytracing techniques to train neural networks \cite{b26,b28}. Despite this innovative approach, the application of such algorithms is hampered by the discrepancy between simulated and empirical data, leading to compromised accuracy.

{
Overall, both model-based and data-driven methods exhibit distinct strengths and limitations in radiomap reconstruction \cite{b30}. Model-based techniques are heavily contingent on predefined models, which may struggle to accurately represent complex real-world scenarios. Conversely, the efficacy of data-driven methods hinges on the availability and integrity of data samples, making them susceptible to significant performance degradation when addressing inadequate data volumes or substandard data quality.
Given these inherent challenges, there is an increasing trend towards integrating model-based and data-driven strategies, which aims to reduce the dependency on empirical models and voluminous datasets, potentially enhancing robustness and adaptability in radiomap reconstruction.} 

Moreover, MB-RMR diverges from the conventional uni-band radiomap estimation, which predominantly concentrates on spatial correlations across distinct locales. The MB-RMR approach is poised to harness not only intra-band spatial correlations but also inter-band correlations at identical spatial coordinates. This dual-correlation exploitation is particularly advantageous for novel applications. For instance, MB-RMR could facilitate the generation of coverage maps in the spectrum of fifth-generation (5G) telecommunication based on the coverage maps of its predecessor, fourth-generation (4G) technology, especially since certain 5G protocols do not convey locational data to base stations.

However, the deployment of learning-based MB-RMR methods in real-world settings is fraught with challenges. These methods necessitate a comprehensive set of structured ground truth data, such as images and third-order tensors, which are essential for the backpropagation process during training \cite{b31,b35}. The sparse nature of actual drive test data further exacerbates the difficulty of implementing these methods effectively for training purposes.

To overcome the outlined obstacles, we introduce a novel framework, \textbf{RadioGAT}, which integrates model-based and data-driven methods within a Graph Attention Network (GAT) for MB-RMR. RadioGAT specifically addresses the issue of dataset disparity by focusing on radiomap interpolation within a singular region, thereby eliminating the discrepancy between training and testing datasets encountered in real-world scenarios. By initially representing the sparse observation as a sparse graph via a channel propagation model, RadioGAT employs GAT to facilitate cross-band radiomap extrapolation, thereby reducing the framework's reliance on extensive data. { Through the integration of model-based and data-driven methods, our proposed RadioGAT has superior performance in accuracy and robustness, particularly under insufficient training samples.}

Our contributions can be summarized as follows:

\begin{itemize}
    \item \textbf{A Joint Model-based and Data-driven Reconstruction Framework for Multi-band Radiomap}: Unlike existing MB-RMR literature, RadioGAT is introduced as a novel framework that synergizes model-based and data-driven approaches to enhance MB-RMR accuracy with limited data. The framework bifurcates the problem into two components: 1) model-based spatial-spectral correlation encoding; and 2) data-driven cross-band generalization. The former's output serves as a graph structure, providing prior information for the latter's input and facilitating node-level mask in practical deployment. 
    \item \textbf{Model-based Spatial-spectral Correlation Encoding}: From the perspective of a graph, observable grids of different frequencies and spaces are modelled as different nodes. A novel frequency-dependent radio depth map is introduced for spatial-spectral correlation encoding in MB-RMR. Such correlation map is further utilized to construct an adjacency matrix in the graph, which simultaneously integrates the information from the urban map and radio propagation model.
    
    \item \textbf{Data-driven Method for Cross-band Radiomap Generalization}: To utilize the data collection in MB-RMR, the observations at different positions and frequencies are input into the modelled graph as node attributes, which enables feature input for nodes. To promote efficient propagation of features between nodes, GAT is applied for cross-band radiomap generalization with the input of node features and adjacency matrix in the graph.

    \item \textbf{Experimental Evaluation for Supervised and Semi-supervised Learning}: Ten real-world environmental maps sourced from OpenStreetMap\footnote{https://www.openstreetmap.org/} serve as the basis for generating radiomap datasets using ray tracing technology across various frequency bands. Empirical results from these datasets confirm the superiority of the proposed spatial-spectral correlation encoding approach. Furthermore, RadioGAT outperforms existing state-of-the-art methods in terms of efficiency and robustness and exhibits strong performance in a semi-supervised learning setting, thus indicating strong deployability under limited data.
\end{itemize}

The remainder of this work is organized as follows:
We first overview the related work in Section \ref{related}, after which we introduce the system model and problem formulation in Section \ref{sysmodel}. Followed by the presentation of model-based correlation encoding methods for RadioGAT in Section \ref{sec: model-based}, we introduce the details of the training process of data-driven radiomap cross-band generalization in Section \ref{sec: training}. Comprehensive experimental results are presented in Section \ref{sec: experiment}.
Finally, we summarized our work in Section \ref{sec: conclusion}.

\section{Related Works} \label{related}

\subsection{Uni-band Radiomap Reconstruction}
Existing works in conventional uni-band RMR usually can be categorized into 1) model-based methods and 2) data-driven methods \cite{b12}. 

\subsubsection{Model-based RMR} Model-based methods usually assume a certain radio propagation for radiomap estimation, such as log-distance path loss model \cite{6150651}. Other typical examples
include thin-plate splines \cite{5934611}, inverse distance weight interpolation \cite{b23},  parallel factor analysis \cite{7175044}, compressed sensing\cite{b19}, tensor decomposition \cite{b20}, radial basis function kernels \cite{b24} and etc. 

\subsubsection{Data-driven RMR} Different from model-based methods, data-driven methods usually do not assume a radio propagation model and explore the data statics of sparse observations for RMR.
Traditional data-driven methods include Gaussian Process Regression  \cite{b21}, matrix completion \cite{b22} and interpolation methods (such as Kriging \cite{b23} and graph signal processing \cite{8066314}). These methods can be used to construct the space radiomap without specific model assumptions, and are easy to implement with limited Root Mean Square Error (RMSE) performance. 

Beyond traditional data-driven approaches, learning-based RMR has recently attracted significant attention.
These learning-based approaches utilize neural networks to explore implicit expressions of available data features and then estimate the entire radiomap. For example, CNN-based methods, such as UNet\cite{b26}, have been proposed for RMR. To leverage the surrounding environmental information, city-building maps have been considered as additional network input. Limited by the effective receptive field, the mask matrix cannot adequately capture the signal propagation model impacted by the buildings. In addition, GAN-based methods\cite{b28} are also proposed. However, a sufficient pre-generated physical dataset is required for GAN, which is usually inaccessible in real scenarios.

Despite many successes, there are certain limitations in either model-based or data-driven methods. Model-based methods cannot precisely capture the radio propagation models affected by shadowing or obstacles in a complex environment. On the other hand, data-driven methods are sensitive to the quantity and quality of observed samples. 
Accordingly, the effective integration of model-based and data-driven methods should be considered. In uni-band RMR, there have been several attempts to integrate model-based and data-driven approaches.
Taking the line-of-sight and non-line-of-sight information into account, a Graph neural network (GNN) based scheme is proposed in \cite{b29}. In addition, the authors of \cite{b30} proposed a 
GAN-based learning framework to explore the radio propagation patterns from global information while emphasising the shadowing effect from local features. 

\subsection{Multi-band Radiomap Reconstruction}

Featured by the spatial-spectral correlations in different locations over multiple frequency bands, the multi-band radiomap 
is different from the uni-band RMR, which only focuses on the spatial features in a single band. To utilize spatial (intra-band) and spectral (inter-band) correlation simultaneously, 
existing works focus on the 
joint spatial-spectral correlation extraction\cite{b31,b32,b33,b34,b35,b36}. 
The concept of multi-band radiomap was first introduced in \cite{b31} and \cite{b32}, where multi-band radiomaps are separated into various single-frequency radiomaps without considering the propagation frequency fading difference. 
In \cite{b33} and \cite{b34}, the authors proposed to jointly implement the spatial-spectral reconstruction based on interpolation approaches, while all the frequencies are treated equivalently ignoring the spectral radio model information. Moreover, these methods further limited performance in multiple transmitter scenarios. 
In addition, generative learning models, such as conditional GAN (cGAN) and autoencorder, are also introduced in MB-RMR \cite{b35} and \cite{b36}. 
However, the impact of obstacles is not considered in the theoretical description and performance evaluation.

In summary, existing MB-RMR works lack simultaneous consideration of multi-band correlation, together with the impact of obstacles. Inspired by the successful experience in spatial RMR, we shall consider utilizing deep learning methods under the integration of model-based and data-driven methods for MB-RMR.

\section{System Model and Problem Fomulation} \label{sysmodel}
\subsection{System Model}
\begin{figure}[t]
\centering
\includegraphics[width=9cm]{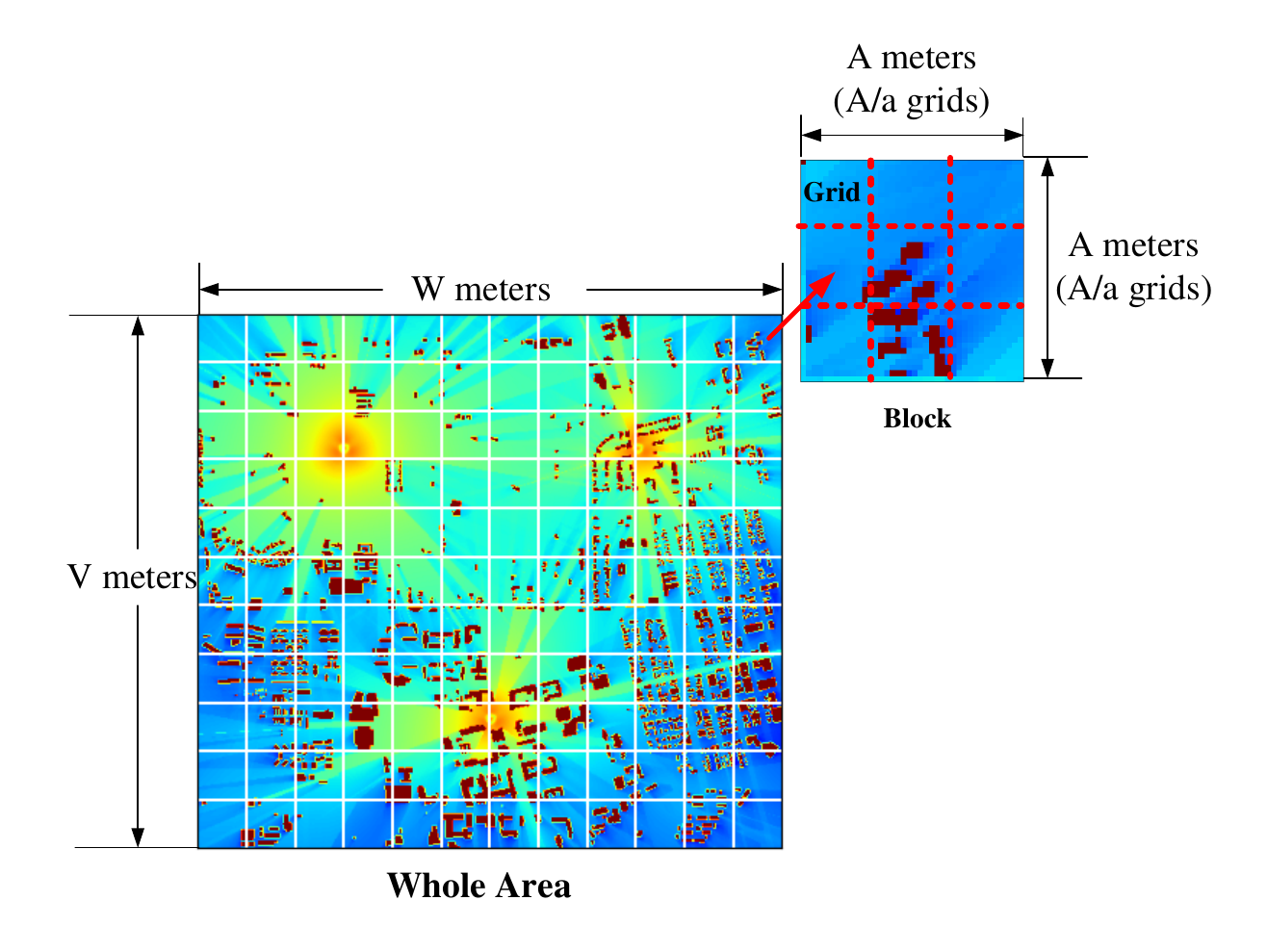}
\caption{The system model under block division at one fixed frequency with the dark red parts for buildings: grid resolution is $a\times a$ meter.}
\label{fig2}
\end{figure}
As shown in Fig. \ref{fig2}, we consider a general area of size $W\times V$ meters for all frequency bands. For generality, suppose that there are $M$ transmitters arbitrarily in the area. Define the transmitter set as $\mathbf{M}=\{1,2,...m,...,M\}$. Moreover, such a given area is equally divided into identical blocks and each block is a small area with the size of $A \times A$ meters. Define the set of blocks as $\mathbf{B}=\{1,...,b,...,N_B\}$. To quantify the observation, each block is further divided into grids (i.e., the pixel of the whole area shown in Fig. \ref{fig2}) with the fixed interval $a$ meters. Define $A/a$ as $l$. For block $b \in \mathbf{B}$, the set of grids is defined as $\mathbf{R_b}=\{r_{b,1},...,r_{b,n},...,r_{b,N}\}$, where $N=l^2$ is the number of grids in block $b$.

\begin{figure}[t]
\centering
\includegraphics[width=5cm]{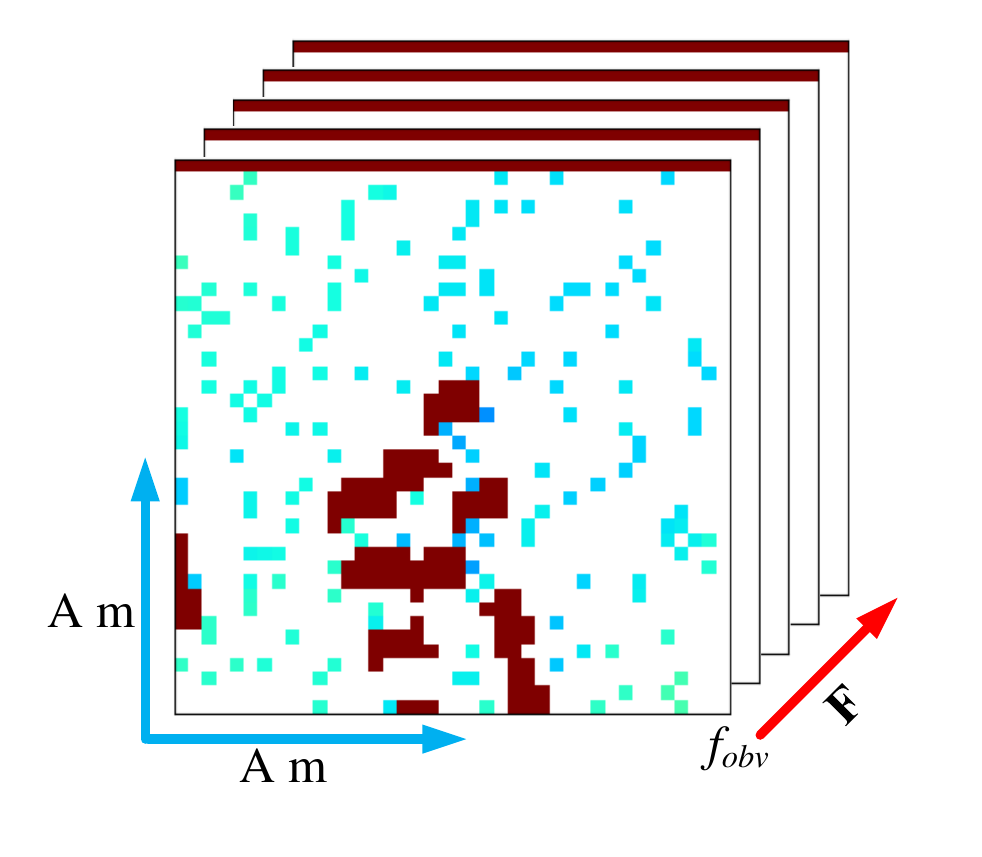}
\caption{The sparse observation block with transmitted signals across K frequencies where the dark red parts indicate the buildings.}
\label{fig3}
\end{figure}

To make the problem more precise, given a block $b\in \mathbf{B}$, the transmitted signals are across $K$ different frequencies for each transmitter. The set of signal frequencies can be denoted by $\mathbf{F}=\{f_{1}, f_{2}, ...f_{k},...f_{K-1}, f_{K}\}$. As shown in Fig. \ref{fig3}, the observation in block $b$ is sparse in both frequency and space. Such sparse observation is described as follows:
{
\begin{itemize}
\item The set of observed grids in block $b$ is defined as $\mathbf{R_{b}^{(obv)}}=\{r_{b,1}^{(obv)},...,r_{b,Nobv}^{(obv)}\}$, in which the number of grids is $N_{obv}$. Note that $N_{obv}$ is usually much smaller than $N$ in realistic applications.

\item For each grid in block $b$, there are $K$ RSS's in $\textbf{F}$. However, only one frequency (i.e., $f_{obv}$) of $\mathbf{F}$ is observed.
\end{itemize}

In practical applications, the sparsity of observations are induced by expensive labour and time costs of multi-band radiomap measurement collection. Next, we will introduce the radio propagation model in detail.}



\subsection{Radio Propagation Model}

According to our model in the previous subsection, the set of grids for block $b$ is $\mathbf{R_{b}}=\{r_{b,1},...,r_{b,n}...,r_{b,N}\}$. Given a grid $r_{b,n}$, the small-scale fading can be rendered negligible by averaging, which facilitates the computation of the average RSS. In this way, the average RSS\cite{b31,b33,b35} at this grid is given by
\begin{equation} \label{radiop}
P^{(k)}_{b,n,m}=P^{(k)}_{m}-L_{c}-\eta_{f} \log _{10}\left(f_{k}\right)-10\eta_{d} \log _{10}\left(d_{b,n,m}\right),
\end{equation}
where $P_{m}^{(k)}$[dBm] is the transmission power of transmitter $m$ over the frequency $f_{k}$, and $L_{c}$[dB] is the constant propagation loss. $\eta_{f}$ and $\eta_{d}$ are the frequency fading factor and distance fading factor respectively. $d_{b,n,m}$ is the distance between grid $r_{b,n}$ and the location of transmitter $m$.

Then, the RSS of the single receiving grid $r_{b,n}$ at frequency $f_k$ can be written as
\begin{equation}
    P^{(k)}_{b,n}=\sum_{m=1}^{M} P^{(k)}_{b,n,m}.
\end{equation}

\subsection{Problem Formulation}
For block $b ,\forall b \in \mathbf{B}$, a complete multi-band radiomap characterizes $P^{(k)}_{b,n}$ ($\forall n \in \{1,...,N\}$ and $k\in \{1,...,K\}$) of each grid. Thus, such radiomap can be defined as

\begin{equation}
p_{b}=\begin{bmatrix}P_{b,1}^{(1)}&\cdots&P_{b,N}^{(1)},\\\vdots&\ddots&\vdots
\\P_{b,1}^{(K)}& \cdots &P_{b,N}^{(K)}\end{bmatrix}
\in\mathbb{R}^{N\times K}.    
\end{equation}

To predict $p_b$,
RSS is observed in $\mathbf{R_b^{(obv)}}$ 
at $f_{obv}$. The observed sparse RSS can be further defined as 
    {$p^{(obv)}_{b,Nobv}=[P_{b,1}^{(obv)},...,P_{b,Nobv}^{(obv)}]$}. 
\begin{figure}[t]
\centering
\includegraphics[width=8cm]{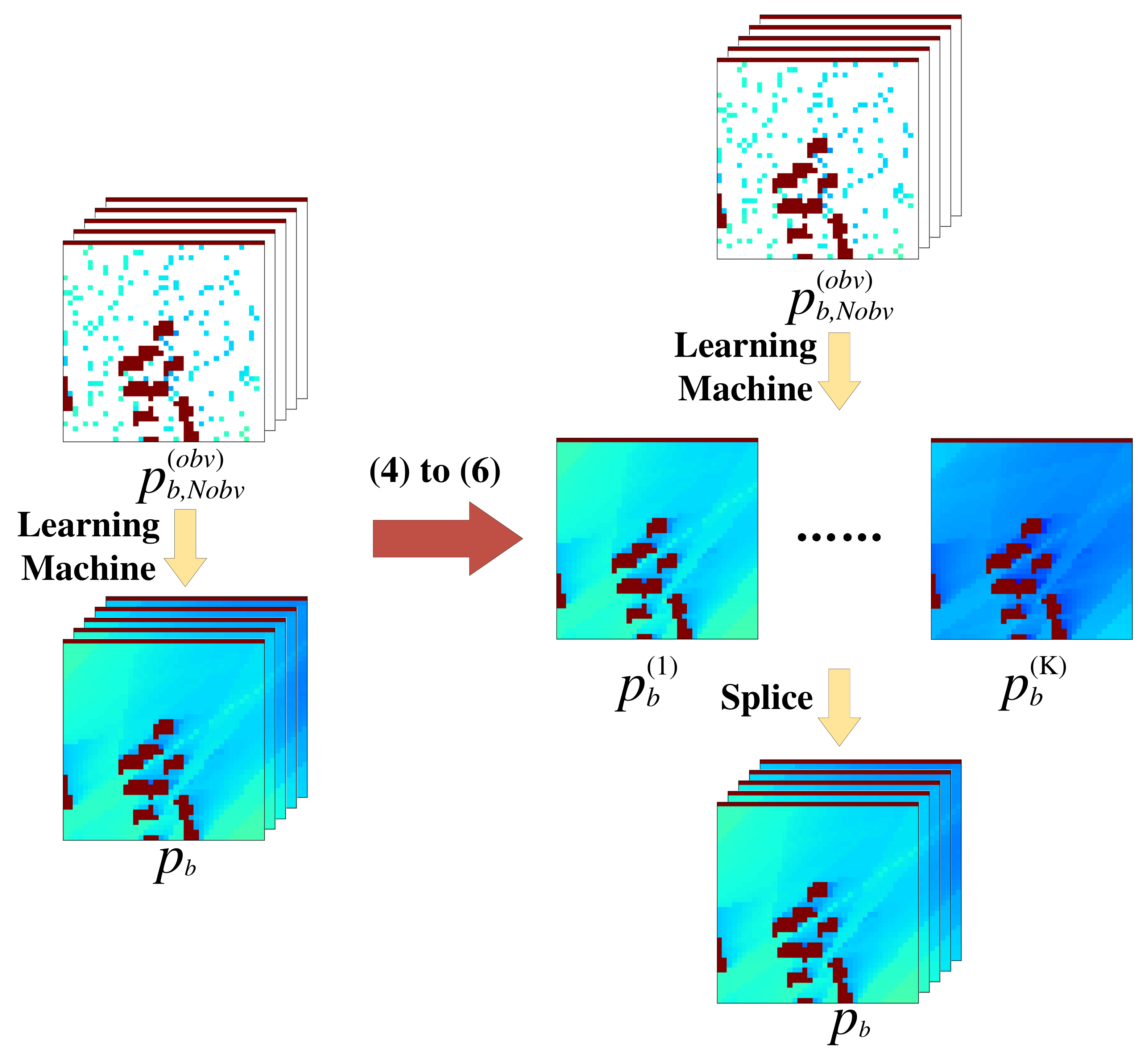}
\caption{The problem transfer of radiomap reconstruction from (4) to (6).}
\label{fig4}
\end{figure}
In this way, the objective of MB-RMR is to find a learning machine $g$ to reconstruct the blank in $p_b$ under the observations in $p^{(obv)}_{b,Nobv}$, i.e.,
\begin{equation}
\hat{p}_b=g(p^{(obv)}_{b,Nobv},f_{obv}),\forall f_{obv}\in \mathbf{F},
\label{fomula 3}
\end{equation}
{where $g(\cdot)$ is the prediction function and $\hat{p}_b$ is a matrix representing the prediction of $p_b$}.
 
As shown in Fig. \ref{fig4}, to reduce the computational complexity, we consider predicting the radiomap of each frequency one by one in block $b$ and finally splicing the prediction results. 
Set $f_{target}\in \mathbf{F}$ as the current target frequency. The radiomap ground truth at $f_{target}$ in block $b$ can be denoted as
\begin{equation}
p_{b}^{(target)}=[P_{b,1}^{(target)},P_{b,2}^{(target)},...,P_{b,N}^{(target)}]\in\mathbb{R}^{N\times 1}.    
\end{equation}

$f_{target}$ can be an input of neural networks to predict the radiomap $p_{b}^{(target)}$. (\ref{fomula 3}) can be rewritten as
\begin{equation}
\hat{p}_b^{(target)}=g(p^{(obv)}_{b,Nobv},f_{obv},f_{target}),\forall f_{obv},f_{target} \in \mathbf{F},
\end{equation}
where $\hat{p}_b^{(target)}$ is the prediction of $p_{b}^{(target)}$. In this way, the problem of MB-RMR for 
finding $g(\cdot)$ can be denoted as
\begin{equation}
\begin{array}{l}
\underset{g}{\min} \left\|g\left(p^{(obv)}_{b,Nobv}, f_{obv}, f_{target }\right)-p_{b}^{(target)}\right\|_{\mathcal{F}} \text{,}\\
\quad \quad \quad\text { s.t. } \quad f_{obv}, f_{target} \in \mathbf{F} \text{,}
\end{array}
\end{equation}
where $||\cdot||_{\mathcal{F}}$ means the Frobenius norm of matrices. {Note that, a global prediction function $g(\cdot)$ is alternating trained by observations in each frequency in this work. In the inference stage, $g(\cdot)$ could predict the radiomap in one target frequency by inputting the corresponding features.}

Above all, we can define the reconstruction process from sparse radiomap $p_{b,Nobv}^{(obv)}$ to completed radiomap $p_{b}^{(target)}$ in block $b\in \mathbf{B}$. $p_b$ can be further spliced by $p_b^{(target)},\forall f_{target} \in \mathbf{F}$. However, simply interpolating $p_{b,Nobv}^{(obv)}$ to obtain $p_b^{(target)}$ without knowing the correlation between different grids and frequencies (i.e., $f_{obv}$ and $f_{target}$) will induct to weak performance. In fact, other information such as the positions of different grids and transmitters, the urban map and the radio propagation model can help reconstruct multi-band radiomaps. To mine the correlations from such information, graphs are considered to evaluate the connections between grids at different frequencies. 
\begin{figure}[t]
\centering
\includegraphics[width=7.5cm]{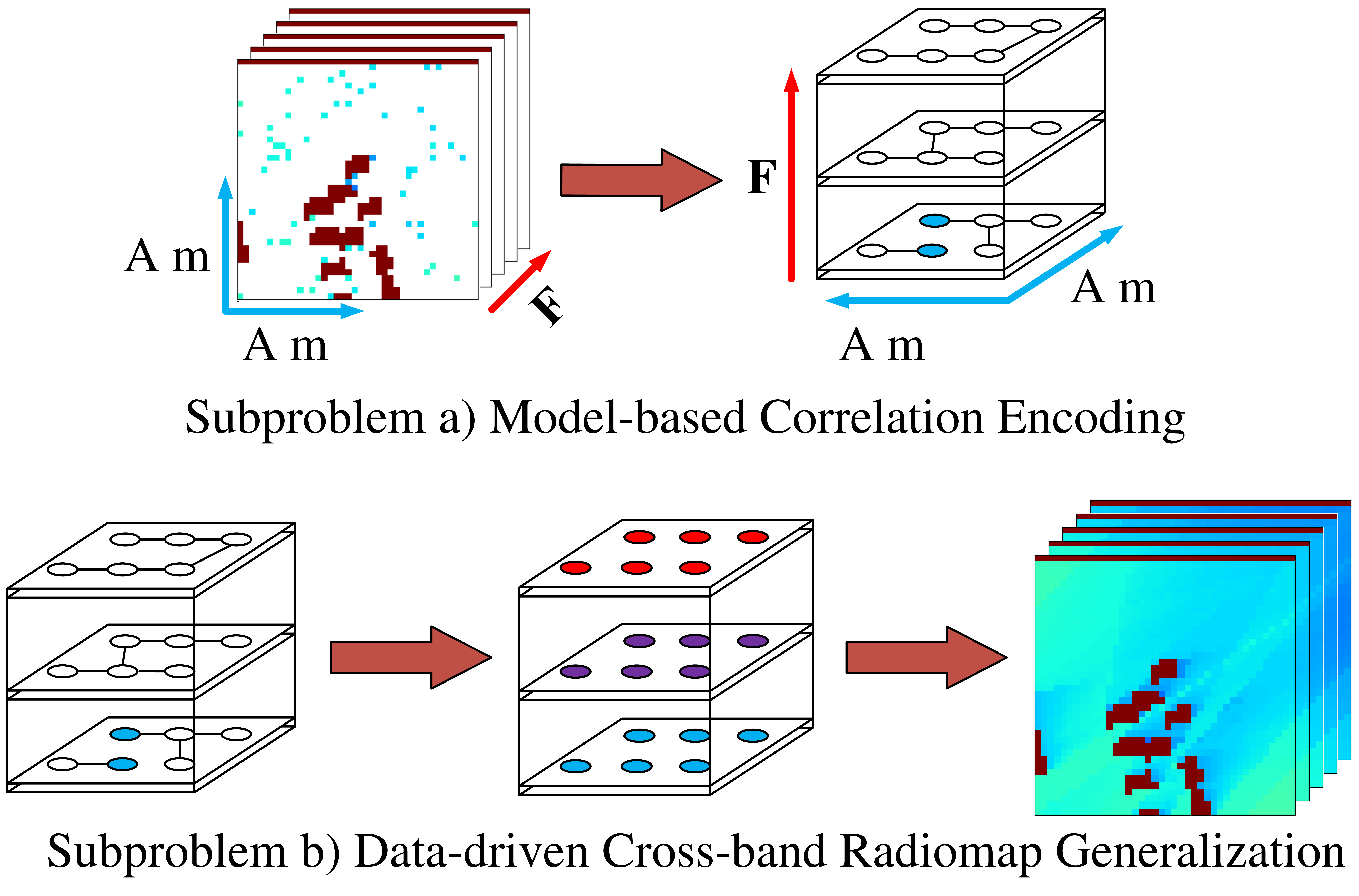}
\caption{The two folds of MB-RMR problem.}
\label{figtwofold}
\end{figure}

From a graph perspective, MB-RMR is decomposed into two subproblems as shown in Fig. \ref{figtwofold}: \textbf{a) model-based correlation encoding} and \textbf{b) data-driven cross-band radiomap generalization}. There is an inheritance relationship between the two subproblems. Specifically, subproblem a) focuses on mining the correlation at different spaces and frequencies without RSS observation, while subproblem b) focuses on implementing cross-band radiomap prediction based on a sparse/small number of observations with the correlations computed in subproblem a). The two subproblems will then be detailly discussed in Section 
\ref{sec: model-based} and Section \ref{sec: training}.
\section{Model-based Correlation Encoding} \label{sec: model-based}
 In this section, we will introduce the model-based correlation encoding method in three folds. First, we model the nodes and edges of a graph under the MB-RMR problem. Then, three main existing node correlation encoding methods for RMR are introduced. Finally, the model-based spatial-spectral correlation encoding method for MB-RMR graph construction is proposed. The performance of different encoding methods will be further analyzed in Section \ref{sec: experiment}.
\subsection{Graph Modelling}
Here, we consider the undirected graph for the radiomap in block $b\in \mathbf{B}$ at $f_{k},\forall f_k \in \mathbf{F}$, i.e., $G^{(k)}_b=(\mathcal{V}^{(k)}_b, \mathcal{E}^{(k)}_b)$. $\mathcal{V}^{(k)}_b$ and $\mathcal{E}^{(k)}_b$ (i.e., the sets of nodes and edges, respectively) of the graph $G^{(k)}_b$ are defined as follows.

\subsubsection{Node} Node $v_{b,n}^{(k)}$ is defined as the observation at $f_{k}$ in grid $r_{b,n}$. The node attribute is the observation of RSS at such node (i.e., $P_{b,n}^{(k)}$ for observed nodes or $0$ for unobserved nodes), whose features include node position, $f_k$ and $f_{target}$.

{For $f_{k}$ of block $b$, the sparse radiomap including observed nodes, unobserved nodes and obstacles can be expressed as a matrix $\mathbf{map}_{b}^{(k)}=(\mathcal{M}_{ij})\in\mathbb{R}^{l\times l}$, where each entry $\mathcal{M}_{ij}$ is calculated by
\begin{equation}
    \mathcal{M}_{ij}= \begin{cases}
   -1\quad &\text{building} \\
0\quad &\text{unobserved RSS} \\
P_{b,n}^{(k)}\quad &\text{observed RSS}\\
\end{cases},
\end{equation}}
Delete $-1$ in $\mathbf{map}_{b}^{(k)}$. Set the remaining elements of $\mathbf{map}_{b}^{(k)}$ attribute to set $\mathbf{RSS}_b^{(k)}$, in which the number of elements (i.e., nodes) is $N_{b}^{(k)}$. Thus $\mathbf{RSS}_b^{(k)}$ can be denoted as 
$\mathbf{RSS}_b^{(k)}=\{P_{b,1}^{(k)},0,...P_{b,n}^{(k)},...,0\}$. $\mathcal{V}_b^{(k)}$ can be further defined as
\begin{equation}
    \mathcal{V}_{b}^{(k)}=\{
v_{b, 1}^{(k)}, \dots, 
v_{b, N_b^{(k)}}^{(k)}\}.
\end{equation}

\subsubsection{Edge} The connection between two nodes $v_{b,i}^{(k)}$ and $v_{b,j}^{(k)}$ is defined as an edge $E \left(v_{b,i}^{(k)},v_{b,j}^{(k)}\right)$. There are two types of edge attributes considered in this paper, connected and unconnected (1 and 0 respectively). If there is a high correlation between two nodes, the edge attribute between the two nodes is connected (1), otherwise the edge attribute between the two nodes is unconnected (0).

Due to the difference in node correlation, in RMR, only nodes with similar attributes (i.e., RSS) should be connected. Usually, a graph $G^{(k)}_b$ can be captured by an adjacency matrix $\mathcal{E}^{(k)}_b$ with each element indicating the edge between two nodes.
Let $E^{(k)}_{b,i,j}$ represent the $(i, j)$-th entry of the graph adjacency matrix $\mathcal{E}_{b}^{(k)}$. It is noted that $E^{(k)}_{b,i,j}$ is equivalent to $E\left(v_{b,i}^{(k)}, v_{b,j}^{(k)}\right)$. {Then, the adjacency matrix is defined by $\mathcal{E}_{b}^{(k)} \in\mathbb{R}^{N_{b}^{(k)} \times N_{b}^{(k)}}$, where each entry $E^{(k)}_{b,i,j}=1$ if connected by an edge; otherwise, $E^{(k)}_{b,i,j}=0$.}

{Note that graph-based radiomap estimation can be formulated as semi-supervised learning \cite{b43}, where RSS can be viewed as annotations of each node while the environmental and geometrical information can be viewed as features. Both annotated and unannotated nodes shall participate in graph learning, where edges are constructed based on features.} Focusing on the construction of $\mathcal{E}_{b}^{(k)}$, we will introduce three existing encoding methods based on adjacency, environmental information, and transmitter information and then propose our novel model-based correlation encoding method. 

\subsection{Existing Node Correlation Encoding Methods}
\begin{figure}[t]
\centering
\includegraphics[width=8.5cm]{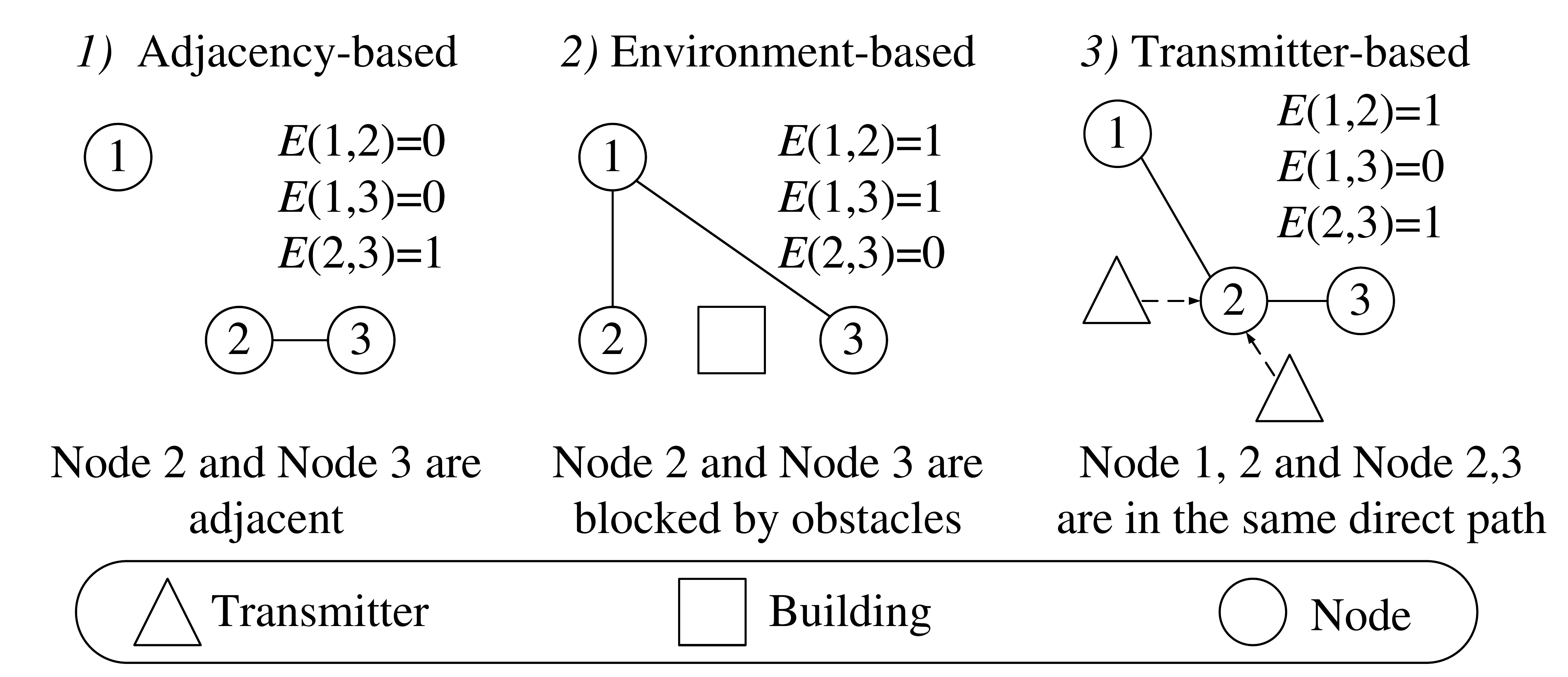}
\caption{{Existing node correlation encoding methods, in which the solid line represents the connection of edges between nodes and the dotted line with an arrow represents the direct path of the base station.}}
\label{graphc}
\end{figure}

As shown in Fig. \ref{graphc}, depending on the perspective of consideration, existing node correlation encoding methods for RMR can be roughly summarized as adjacency-based, environment-based\cite{b29} and transmitter-based methods\cite{b44}.

\subsubsection{Adjacency-based Encoding} According to the radio propagation model, the RSS is more correlated to nearby observations than those far apart\cite{b38}. A straightforward method is to connect adjacent nodes. Then, an intuitive definition of edge in the graph $G^{(k)}_b$ is
\begin{equation}
    E \left(v_{b,i}^{(k)}, v_{b,j}^{(k)}\right)=\left\{\begin{array}{l}
1\text{,} \quad \text { if }  v_{b,i}^{(k)} \text{ and } v_{b,j}^{(k)} \text{ are adjacent;}  \\
0\text{,}\quad \text { otherwise.}
\end{array}\right.
\end{equation}

\subsubsection{Environment-based Encoding}
Since radiomaps in multiple bands rely on the same environmental surroundings, geometric urban information can be used to construct a graph. Taking the urban map as an example, if there are no obstacles between two grid nodes, we can construct an edge between them.
The obstacle information between $v_{b,i}^{(k)}$ and $v_{b,j}^{(k)}$ can be defined as
\begin{equation}
\rho_{b,i,j}^{(k)}=\left\{\begin{array}{l}
1\text{,} \quad \text{if buildings exist between $v_{b,i}^{(k)}$ and $v_{b,j}^{(k)}$\text{;}} \\
0\text{,} \quad \text{otherwise.}
\end{array}\right.
\end{equation}
Therefore, the environment-based edge connection criterion can be defined as

\begin{equation}
    E \left(v_{b,i}^{(k)}, v_{b,j}^{(k)}\right)=\left\{\begin{array}{l}
1\text{,} \quad\text { if } \rho_{b,i,j}^{(k)}=1\text{;}
\\
0\text{,}\quad \text { otherwise. }
\end{array}\right.
\end{equation}

\subsubsection{Transmitter-based Encoding}
The transmitter information can be also used to construct a graph. Taking the transmitter position as an example, if two nodes are in the same direct path of a transmitter, it is considered that there is a connection relationship between them.
Define the positional relationship between $v_{b,i}^{(k)}$ and $v_{b,j}^{(k)}$ and the transmitters $\mathbf{M}$ as 
\begin{small}
\begin{equation}
\gamma_{b,i,j}^{(k)}=\left\{\begin{array}{l}
1\text{,}  \text{ if } v_{b,i}^{(k)},\text{ } v_{b,j}^{(k)} \text{ and } m \text{ are on a straight line, } \exists m\in\mathbf{M}\text{;} \\
0\text{,}  \text{ otherwise.}
\end{array}\right.
\end{equation}
\end{small}
Then the transmitter-based connection criterion can be defined as
\begin{equation}
    E \left(v_{b,i}^{(k)}, v_{b,j}^{(k)}\right)=\left\{\begin{array}{l}
1\text{,} \quad \text { if }  \gamma_{b,i,j}^{(k)}=1\text{;}  \\
0\text{,} \quad \text { otherwise. }
\end{array}\right.
\end{equation}

Current methods for encoding correlations do not adequately incorporate channel models and are limited to considering a singular dimension influencing signal propagation. Subsequently, we will delineate the proposed model-based spatial-spectral correlation encoding technique to more accurately represent the correlation between nodes.

\subsection{Model-based spatial-spectral Correlation Encoding}
According to \cite{b38}, the RSS of the measurement grid depends on the length of the signal through the building of a single transmitter $m$ to the measurement grid. To embed the radio propagation model, we propose $T_{b,n,m}$ to capture the shadowing and fading from obstacles, similar to \cite{zhang2023radiomap}. For each known topological map, the connection from measurement grid $r_{b,n}$ to the transmitter $m$ is defined as $l_{b,n,m}$. Then $T_{b,n,m}$ is denoted as the fraction of non-buildings length in $l_{b,n,m}$.
Thus for a single transmitter $m$, the radio depth value of the node $v_{b,n}^{(k)}$ can be expressed as
\begin{equation}
D_{b,n,m}^{(k)}=T_{b,n,m}(C-\alpha \log_{10}d_{b,n,m}-\beta \log_{10}(f_{k})),
\label{fomula14}
\end{equation}
where $C$ is a hyperparameter reflecting constant fading $P_{m}^{(k)}-L_{c}$, $\alpha$ is a hyperparameter reflecting the distance-decay effect, and $\beta$ is a hyperparameter reflecting the frequency-decay effect. This depth term reflects the decay of PSD considering the obstacles, distance and frequency with respect to Eq. (\ref{radiop}).

{The similarity between
nodes $v_{b,i}^{(k)}$ and $v_{b,j}^{(k)}$ can be characterized by $D_{b,i,m}^{(k)}$ and $D_{b,j,m}^{(k)}$.
Inspired by adjacency-based encoding, the radio propagation characteristics are usually similar in a certain range. 
Limited by a predefined distance threshold $d_{th}$, the distance between nodes shall be smaller than that between the node and the transmitter. Thus, $T_{b,n,m}(C-\alpha \log_{10}d_{b,n,m})$ can be approximated by constant $C_0$ under the constraint of $d_{th}$.
}
Then the radio depth can be approximated by
\begin{equation}
D_{b,n,m}^{(k)}\approx C_0-\beta T_{b,n,m} \log_{10}(f_{k}),
\label{fomula1app}
\end{equation}
where the key pattern of PSD distribution can be captured by an alternative depth term as
\begin{equation}
\hat{D}_{b,n,m}^{(k)}=\beta T_{b,n,m} \log_{10}(f_{k}),
\end{equation}
with $\beta$ serving as a hyperparameter to adjust the range of the desired depth feature.

For convenience, we shall use the simplified alternative depth $\hat{D}_{b,n,m}^{(k)}$ to capture the texture patterns in the radio depth map.
Then, for multiple simultaneous transmitters $\{1,...,M\}$, the signal depth value of the node $v_{b,n}^{(k)}$ in block $b$ can be expressed as
\begin{equation}
\hat{D}_{b,n}^{(k)}=\sum_{m=1}^{M}\hat{D}_{b,n,m}^{(k)}.
\end{equation}

\begin{figure}[t]
\centering
\includegraphics[width=8.8cm]{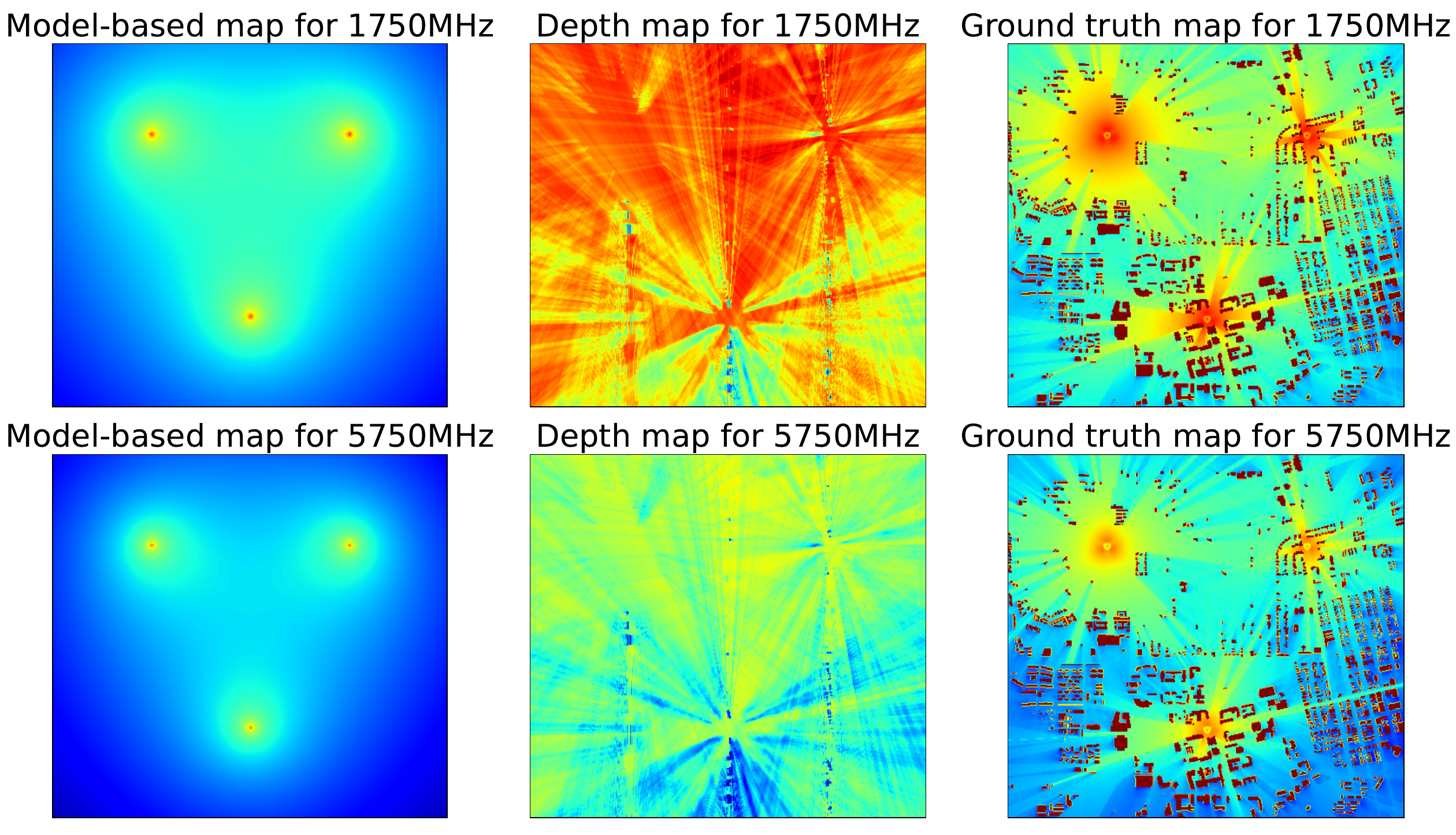}
\caption{The visual comparison among model-based coverage map, depth map and ground truth radiomap.}
\label{fig6}
\end{figure}

{In practice, the radio depth of an area can be calculated block by block. To illustrate the difference between a radio depth map and a traditional model-based coverage map, we also show the model-based coverage based on the log-distance model [24]. As shown in Fig. \ref{fig6}, the computed depth map in the area can better reflect the impact of building occlusion compared to the model-based coverage map. The distance factor can be captured through the position feature of the node and the distance threshold $d_{th}$.}

With the defined $\hat{D}_{b,n}^{(k)}$, the correlation between two nodes $v_{b,i}^{(k)}$ and $v_{b,j}^{(k)}$ can be characterized by
\begin{equation}
\Delta \hat{D}_{b,i,j}^{(k)}=\left |\hat{D}_{b,i}^{(k)}-\hat{D}_{b,j}^{(k)}\right | .
\end{equation}

Since the depth value can roughly reflect the model characteristics of the received signal, the depth difference can capture the correlation between different nodes. Set $\delta$ as the depth connection threshold. Here, { $d_{th}$ and $\delta$ are hyperparameters that control the sparsity of the graph, where $d_{th}$ controls the sparsity affected by distance while $\delta$ controls the sparsity affected by frequency and occlusion.}
The depth-based edge connection criterion can be expressed as
\begin{equation}
    E \left(v_{b,i}^{(k)}, v_{b,j}^{(k)}\right)=\left\{\begin{array}{l}
1 \text {, if }d_{b,i, j} \leq d_{th}\text { and }\left | \Delta \hat{D}_{b,i,j}^{(k)} \right | \leq \delta \text{;}  \\
0\text {, otherwise;}
\end{array}\right.
\end{equation}
where $d_{b,i,j}$ is the distance between grid $r_{b,i}$ and $r_{b,j}$.

According to the above procedure, we have constructed the graph at $f_{k}$. In reality, the signal fading varies at different frequencies. To characterize fading at different frequencies, the node correlation encoding procedure should be executed in all frequencies of $\mathbf{F}$. 

{\textbf{Discussion:}
In this work, we present a model-based correlation encoding approach capable of mapping the interplay between transmitters and grids, together with the inter-grid relationships. Our RadioGAT framework is designed with flexibility to be integrated with different edge encoding methods to meet diverse contextual requirements.
Specifically, adjacency-based encoding can be applied in RadioGAT in scenario with the absence of environmental and station data. If only environmental data is available, environment-based encoding shall lead to better performance in RadioGAT. Moreover, transmitter-based encoding is more suitable with exclusive base station information. The performance implications of these integrated encoding strategies within RadioGAT will be further discussed in Section \ref{sec: experiment}.}

\section{Data-driven Cross-band Radiomap Generalization} \label{sec: training}
\begin{figure*}[t]
\centering
\includegraphics[width=15cm]{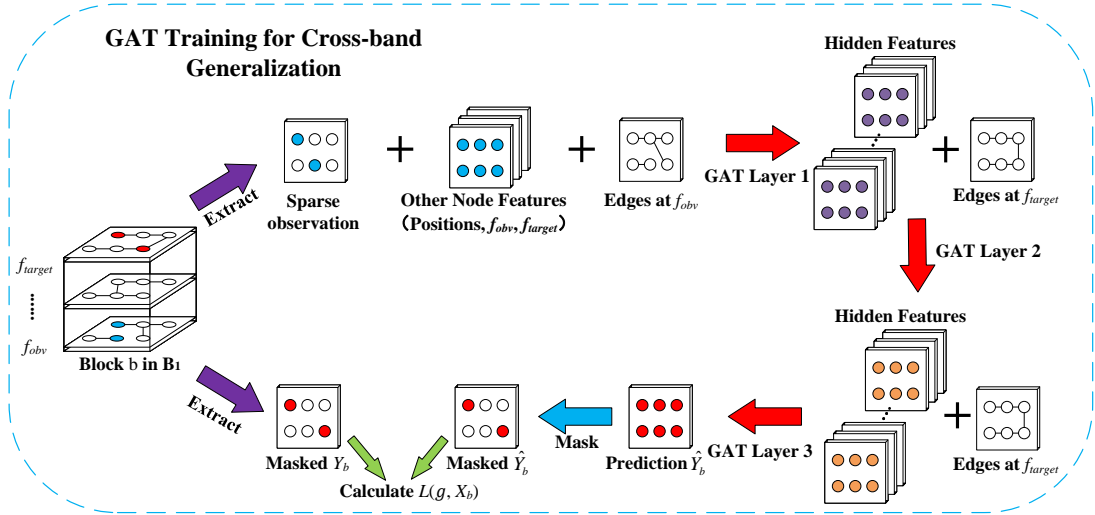}
\caption{The training process of GAT for cross-band generalization.}
\label{fig7}
\end{figure*}
With the constructed graph, GAT is introduced to complete cross-band radiomap generalization under sparse observation. Since the observable areas of different blocks are not exactly the same, the number of nodes in each block is different. Here, we apply the inductive GAT \cite{b43} as the backbone to learn the weights between connected nodes and predict the signal strength of uncollected nodes. 

For the task of estimating $p_{b}^{(target)}$ from $p^{(obv)}_{b,Nobv}$, necessary measurement of $p_{b}^{(target)}$ is conducted in some blocks of the whole area. Set these measured blocks belong to set $\mathbf{B_1}$, while the other blocks belong to set $\mathbf{B_2}$. $\mathbf{B_1}$ and $\mathbf{B_2}$ are utilized for the training and inference procedure of RadioGAT, respectively. We will next introduce the training procedure in this section and then introduce the inference procedure in Section \ref{sec: experiment}. As shown in Fig. \ref{fig7}, the training procedure of RadioGAT consists of three parts: 1) network input initialization; 2) network structure and output; and 3) masked GAT for supervised and semi-supervised learning. 
\subsection{Network Input Initialization}
The network input consists of two parts: input features and graph adjacency matrix. 

\subsubsection{Graph Node Features} The input features consist of four channels. Specifially, $p^{(obv)}_{b,Nobv}$ are extracted from block $b, \forall b \in \mathbf{B_1}$ as a channel. Other node features, including the positions of nodes, $f_{obv}$ and $f_{target}$, are the input of the neural network, which mainly captures the information of distance and frequency correlation. 
The sparse observation and node features at node $v_{b,n}^{(obv)}$ are denoted as $x_{b,n}$. Then, for block $b$, the GAT input can be expressed by $X_{b}=\{x_{b,1},...,x_{b,n},...,x_{b,N_{b}^{(obv)}}\}$. Before being fed into the GAT, all channels of the input features are normalized.

\subsubsection{Graph Adjacency Matrix} The graph adjacency matrix is the constructed $\mathcal{E}_b^{(k)}$ at $f_{obv}$, i.e., $\mathcal{E}_b^{(obv)}$, which captures the information of the radio propagation and the message passing in GAT. { The geometric information is also embedded in the graph construction. Here the positions of the grid nodes in the entire area of interest are used as input features, rather than the relative position in each block. In addition, the radio depth map is applied in graph construction, where the depth value depends on the information of the entire area. This allows the subsequent network to partially consider the inter-block correlations while processing the correlations between grids within a block.}

\subsection{Network Structure and Output} \label{networkstruct}

The GAT network is comprised of three GAT layers. The output of each GAT layer is calculated by aggregating the input representations of the nodes.
Based on the feature input $X_b$ and graph adjacency matrix $\mathcal{E}_b^{(obv)}$, the GAT layer 1 transfers the discrete observation of $f_{obv}$ to that of $f_{target}$. Then, the remaining GAT layers set the output of the previous GAT layer as feature input and the adjacency matrix at $f_{target}$, i.e., 
 $\mathcal{E}_b^{(target)}$ as graph adjacency matrix input to complete the feature propagation from partially annotated nodes to all nodes at $f_{target}$. Finally, the last GAT layer outputs the RSS prediction $\hat{Y}_{b}$ at each node. $\hat{Y}_{b}$ is defined by $\hat{Y}_{b}=\{\hat{y}_{b,1},...,\hat{y}_{b,n},...,\hat{y}_{b,N_{b}^{(target)}}\}$, where $\hat{y}_{b,n}$ is the prediction of node $v_{b,n}^{(target)}$.

\subsection{Masked GAT for Supervised and Semi-supervised Learning}

Traditional learning-based RMR methods have conventionally transformed grid elements into structured data forms like images or tensors, requiring comprehensive ground truth datasets for each grid element during training.

However, obtaining such complete datasets is typically unfeasible, presenting a notable challenge. Masked training, as detailed in \cite{b42}, offers a solution by selectively hiding outputs during training, advantageous for semi-supervised learning with limited labels. Structured data-based methods, however, lack the flexibility to implement this strategy.

RadioGAT, in contrast, proposes a graph-based architecture inherently compatible with masked training, allowing selective ground truth disclosure at the node level. This design supports both supervised and semi-supervised learning, depending on ground truth data availability, and is pivotal for reconstructing radiomaps from sparsely sampled data. RadioGAT thus provides a resilient alternative to traditional methods, enabling accurate radiomap predictions even with limited ground truth. Further details on node-level masking in RadioGAT will be provided as follows.

\begin{table*}[t]
\centering
\caption{Dataset Parameters}
\begin{tabular}{|c|c|c|c|c|c|c|c|c|}
\hline
Area Index & Longitude & Latitude  & Length (m) & Width (m) & Grid size ($\mathrm{m^2}$) & Transmitter 1 & Transmitter 2 & Transmitter 3 \\ \hline
1          & 38.5472    & -121.7542 & 2400       & 2200      & 5×5             & (121 111)     & (363 111)     & (242 334)     \\ \hline
2          & 35.0291    & -80.7075  & 2800       & 2000      & 5×5             & (143 105)     & (428 105)     & (286 314)     \\ \hline
3          & 34.968     & -80.7685  & 2600       & 2000      & 5×5             & (133 108)     & (398 108)     & (266 325)     \\ \hline
4          & 30.6657    & -88.0831  & 2200       & 2600      & 5×5             & (119 130)     & (356 130)     & (237 391)     \\ \hline
5          & 42.6484    & -71.6345  & 2800       & 2200      & 5×5             & (144 117)     & (431 117)     & (288 351)     \\ \hline
6          & 32.7694    & -97.8055  & 2400       & 2400      & 5×5             & (127 120)     & (380 120)     & (253 360)     \\ \hline
7          & 41.6958    & -88.3437  & 2600       & 2600      & 5×5             & (130 131)     & (390 131)     & (260 393)     \\ \hline
8          & 41.1402    & -104.7811 & 2800       & 1800      & 5×5             & (146 97)      & (439 97)      & (293 290)     \\ \hline
9          & 41.1267    & -104.8027 & 2400       & 2000      & 5×5             & (125 105)     & (376 105)     & (251 316)     \\ \hline
10         & 41.1891    & 104.8042  & 2000       & 2400      & 5×5             & (109 124)     & (327 124)     & (218 372)     \\ \hline
\end{tabular}
\label{table1}
\end{table*}

\begin{table}[t]

\caption{\centering{Algorithm RSS Input Contrast}} 

\label{table2}
\centering
\scalebox{0.95}{
\begin{tabular}{|c|c|}
\hline

Algorithm    & RSS Input Information        \\ \hline

RadioGAT & \begin{tabular}[c]{@{}c@{}}$p_{b,Nobv}^{(1750)},\forall b \in \mathbf{B_2}$ \\ $p_{b,Nobv}^{(1750)}$ and $p_{b,N}^{(target)},\forall b \in \mathbf{B_1}$\end{tabular}  \\ \hline
GNN & \begin{tabular}[c]{@{}c@{}}$p_{b,Nobv}^{(1750)},\forall b \in \mathbf{B_2}$ \\ $p_{b,Nobv}^{(1750)}$ and $p_{b,N}^{(target)},\forall b \in \mathbf{B_1}$\end{tabular}  \\ \hline
Autoencoder  & \begin{tabular}[c]{@{}c@{}}$p_{b,Nobv}^{(1750)}$, $p_{b,Nobv}^{(2750)}$ and $p_{b,Nobv}^{(3750)},\forall b \in \mathbf{B_2}$ \\$p_{b,Nobv}^{(1750)}$, $p_{b,Nobv}^{(2750)}$, $p_{b,Nobv}^{(3750)}$ and $p_{b,N}^{(target)}\forall b \in \mathbf{B_1}$\end{tabular}   \\ \hline
CGAN  & \begin{tabular}[c]{@{}c@{}}$p_{b,Nobv}^{(1750)}$, $p_{b,Nobv}^{(2750)}$ and $p_{b,Nobv}^{(3750)},\forall b \in \mathbf{B_2}$ \\$p_{b,N}^{(1750)}$, $p_{b,N}^{(2750)}$, $p_{b,N}^{(3750)}$ and $p_{b,N}^{(target)}\forall b \in \mathbf{B_1}$\end{tabular}   \\ \hline
SF-Kriging      & \begin{tabular}[c]{@{}c@{}}$p_{b,Nobv}^{(1750)},\forall b \in \mathbf{B_2}$\\ $p_{b,N}^{(1750)}$ and $p_{b,N}^{(target)},\forall b \in \mathbf{B_1}$\end{tabular} \\ \hline
3D-IDW          & \begin{tabular}[c]{@{}c@{}}$p_{b,Nobv}^{(1750)}$, $p_{b,Nobv}^{(2750)}$ and $p_{b,Nobv}^{(3750)},\forall b \in \mathbf{B_2}$\end{tabular}  \\ \hline
TensorCompletion          & \begin{tabular}[c]{@{}c@{}}$p_{b,Nobv}^{(1750)}$, $p_{b,Nobv}^{(2750)}$, $p_{b,Nobv}^{(3750)}$ and $p_{b,Nobv}^{(b)},\forall b \in \mathbf{B_2}$\end{tabular}  \\ \hline
\end{tabular}}
\end{table}

{As defined in Section \ref{networkstruct}, $\hat{Y}_b$ consists all prediction of nodes $\mathcal{V}_b^{(target)}$ at $f_{target}$. However, there are only a few nodes with ground truth observation at $f_{target}$. Thus, masked GAT training is applied. Define nodes with annotated observations at $f_{target}$ as masked nodes, and $z_{b,n}$ as the node indicator function, where $z_{b,n}=1$ indicates that the observation at $v_{b,n}^{(target)}$ is known; otherwise $z_{b,n}=0$. Only the predicted values of the masked nodes contribute to the computation of the loss function for backpropagation. Denote the loss function of neural network $g$ during masked GAT training by  $L_g(g,X_{b},\mathcal{E}_b^{(obv)},\mathcal{E}_b^{(target)})$. It can be calculated as 
\begin{equation}   L_g=\frac{1}{N_{b}^{(target)}} \sum_{n=1}^{N_{b}^{(target)}} z_{b,n}\left(y_{b,n}-\hat{y}_{b,n}\right)^{2}, \forall b \in \mathbf{B_1},
\end{equation}
where $y_{b,n}$ is the normalized result of the RSS ground truth at node $v_{b,n}^{(target)}$.}
Only the masked nodes are subject to penalization by the loss function, thereby enabling the application of semi-supervised learning within the RadioGAT framework. This approach necessitates merely a sparse or minimal set of labelled measurements in block $b$, significantly reducing the cost associated with realistic measurements. All blocks encompassed in 
$\mathbf{B_1}$ are input into GAT as samples with the objective of minimizing $L$. The model $g$ undergoes optimization via the Adam algorithm throughout the GAT training phase. 



\section{Experiment Results} \label{sec: experiment}
In this section, we present the experiment setup and performance compared to existing approaches.
\subsection{Dataset}

To evaluate the performance of the proposed scheme, we use the radiomaps generated from real environments based on ray tracing techniques (a subset of dataset\footnote{https://github.com/BRATLab-UCD/Radiomap-Data}). Specifically, OpenStreetMap is used to obtain real-world building maps of ten areas. The parameters of all areas are listed in Table \ref{table1}. The centre location of each area is represented by latitude and longitude. The size of each area is reflected by the length and width of the area. Each area is divided into multiple grids with the size of $5 \times 5 \mathrm{m^2}$. Then the building maps of all areas are imported into WinProp \footnote{https://web.altair.com/winprop-telecom}. In each area, there are $3$ transmitters operating at five frequencies $\{1750,2750,3750,4750,5750\}$ (MHz) simultaneously. We set the first grid coordinate in the northwest corner of the area as (1, 1), and the east and south as positive coordinates. Then, the positions of transmitters in each area can be represented as shown in Table \ref{table1}. Under these parameter settings, the corresponding radiomap considering the given building map is generated using ray tracing technology. 

\subsection{Experimental Setup}

\begin{table*}[t]
\captionsetup{labelfont={blue},textfont={blue}}
\caption{RMSE ($\rm{dB}$) Contrast under Different Correlation Encoding Methods ($N_{obv}/N=5\%$)}
\centering
\label{table3}
\scalebox{1}{
\begin{tabular}{cccccc}
\toprule
Area                     & $f_{target}$/MHz & Adjacency-based & Environment-based & Transmitter-based & Model-based  \\ \midrule
\multirow{2}{*}{1}       & 4750  & 3.8135    & 3.3301      & 3.7221      & \textbf{2.6655} \\ 
                         & 5750  & 4.5667    & 3.2866      & 4.3223      & \textbf{3.0047} \\ 
\multirow{2}{*}{2}       & 4750  & 3.068     & 3.4157      & 3.2493      & \textbf{3.0924} \\ 
                         & 5750  & \textbf{3.0529}    & 3.3975      & 3.4346      & 3.1522 \\ 
\multirow{2}{*}{3}       & 4750  & 3.0473    & 3.1781      & 3.128       & \textbf{2.8217} \\ 
                         & 5750  & 3.1825    & 2.9978      & 3.3717      & \textbf{2.9581} \\ 
\multirow{2}{*}{4}       & 4750  & 3.3868    & 3.3899      & 3.9508      & \textbf{2.6096} \\  
                         & 5750  & 3.4174    & 3.2858      & 3.5412      & \textbf{2.5671} \\
\multirow{2}{*}{5}       & 4750  & 3.3839    & 3.3079      & \textbf{3.1576}      & 3.2834 \\ 
                         & 5750  & 3.3545    & 3.0238      & 4.1584      & \textbf{2.9012} \\ 
\multirow{2}{*}{6}       & 4750  & 2.8355    & 2.7729      & 3.2745      & \textbf{2.4062} \\ 
                         & 5750  & 4.4512    & 3.0232      & 3.0972      & \textbf{2.4659} \\ 
\multirow{2}{*}{7}       & 4750  & 2.87      & 2.2388      & 3.1323      & \textbf{2.1593} \\ 
                         & 5750  & 3.0362    & 3.3911      & 3.2903      & \textbf{2.2763} \\ 
\multirow{2}{*}{8}       & 4750  & 4.2971    & 4.2306      & 4.2509      & \textbf{3.4609} \\ 
                         & 5750  & 4.0573    & 3.8885      & 4.0082      & \textbf{3.3367} \\ 
\multirow{2}{*}{9}       & 4750  & 4.3703    & 3.7217      & 5.2997      & \textbf{3.145}  \\ 
                         & 5750  & \textbf{3.6583}    & 3.7714      & 4.8398      & 4.0472 \\ 
\multirow{2}{*}{10}      & 4750  & 3.9861    & 3.2442      & 3.592       & \textbf{2.7396} \\ 
                         & 5750  & 3.7526    & 4.1391      & 3.4654      & \textbf{3.2218} \\ 
\multirow{2}{*}{Average} & 4750  & 3.5059    & 3.283       & 3.6758      & \textbf{2.8384} \\ 
                         & 5750  & 3.653     & 3.4205      & 3.7529      & \textbf{2.9931} \\ \bottomrule
\end{tabular}}
\end{table*}

\begin{table*}[t]
\captionsetup{labelfont={blue},textfont={blue}}
\caption{RMSE ($\rm{dB}$) Contrast under Different Algorithms ($N_{obv}/N=5\%$)}
\centering
\label{table4}
\begin{tabular}{ccccccccc}
\toprule
Area                     & $f_{target}$/MHz & RadioGAT & GNN    & Autoencoder & CGAN    & SF-Kriging & 3D-IDW & TensorCompletion \\ \midrule
\multirow{2}{*}{1}       & 4750  & \textbf{2.6655}  & 3.758  & 8.2419      & 12.3569 & 4.118      & 5.0708 & 38.2020          \\ 
                         & 5750  & \textbf{3.0047}  & 4.1918 & 8.2504      & 12.6726 & 4.162      & 6.6781 & 38.9797          \\ 
\multirow{2}{*}{2}       & 4750  & \textbf{3.0924}  & 3.3303 & 6.6907      & 14.6902 & 3.9364     & 4.9361 & 38.6641          \\  
                         & 5750  & \textbf{3.1522}  & 3.4034 & 6.9461      & 11.5335 & 4.0106     & 6.4929 & 39.7977          \\ 
\multirow{2}{*}{3}       & 4750  & \textbf{2.8217}  & 3.0992 & 13.5340     & 13.9071 & 3.6879     & 4.815  & 41.1186          \\ 
                         & 5750  & 2.9581  & \textbf{2.8574} & 14.3369     & 13.2193 & 3.7066     & 6.4656 & 42.0631          \\ 
\multirow{2}{*}{4}       & 4750  & \textbf{2.6096}  & 3.2047 & 7.5784      & 12.8028 & 3.767      & 4.9744 & 37.7393          \\ 
                         & 5750  & \textbf{2.5671}  & 3.0226 & 7.4549      & 12.3885 & 3.8213     & 6.6377 & 38.9156          \\ 
\multirow{2}{*}{5}       & 4750  & 3.2834  & 2.9828 & 7.3149      & 10.1915 & 3.5203     & 4.7769 & 37.8504          \\ 
                         & 5750  &\textbf{2.9012}  & 3.1966 & 7.3939      & 12.6006 & 3.6303     & 6.3829 & 38.9153          \\ 
\multirow{2}{*}{6}       & 4750  & \textbf{2.4062}  & 3.4545 & 7.4560      & 11.2629 & 3.4834     & 4.5184 & 38.7486          \\  
                         & 5750  & \textbf{2.4659}  & 2.7801 & 7.4814      & 13.7752 & 3.5862     & 6.0974 & 98.3507          \\ 
\multirow{2}{*}{7}       & 4750  & \textbf{2.1593}  & 3.1255 & 6.3004      & 10.0153 & 2.9094     & 4.3822 & 39.9987          \\ 
                         & 5750  & \textbf{2.2763}  & 2.5273 & 6.3479      & 11.9420 & 2.983      & 6.064  & 41.2197          \\ 
\multirow{2}{*}{8}       & 4750  & \textbf{3.4609}  & 4.4873 & 13.7668     & 16.5935 & 4.3406     & 5.4446 & 41.1886          \\ 
                         & 5750  & \textbf{3.3367}  & 3.6906 & 14.3320     & 15.7780 & 4.367      & 7.0287 & 42.1145          \\
\multirow{2}{*}{9}       & 4750  & \textbf{3.145}   & 3.1712 & 7.3002      & 11.2742 & 4.1359     & 5.2549 & 41.2579          \\ 
                         & 5750  & 4.0472  & \textbf{3.6153} & 7.1845      & 12.3076 & 4.255      & 6.9063 & 42.3595          \\
\multirow{2}{*}{10}      & 4750  & \textbf{2.7396}  & 3.7347 & 7.9314      & 13.7200 & 3.9736     & 5.2633 & 37.0471          \\  
                         & 5750  & \textbf{3.2218}  & 3.3036 & 8.0106      & 13.3131 & 3.9941     & 6.9216 & 38.1168          \\
\multirow{2}{*}{Average} & 4750  & \textbf{2.8384}  & 3.4348 & 8.6115      & 12.6814 & 3.7873     & 4.9437 & 39.1815          \\
                         & 5750  & \textbf{2.9931}  & 3.2589 & 8.7739      & 12.9530 & 3.8516     & 6.5675 & 46.0833          \\ \bottomrule
\end{tabular}
\end{table*}

In the experiment, we compare our algorithm to the state-of-the-art (SOTA) algorithms, including GNN\cite{b29}, autoencoder\cite{b36}, cGAN\cite{b36}, SF-Kriging \cite{b34}, 3D-IDW\cite{b40}, and TensorCompeltion\cite{b41}. To supplement the explanation, we use the same network structure and hyperparameters as \cite{b29,b36} when benchmarking. For ease of explanation, $p^{(k)}_{b,Nobv}$ represents a random sparse observation at $f_{k}$ in block $b$, and the number of observations is $N_{obv}$. $p^{(k)}_{b,N}$ represents completed observation at $f_{k}$ in block $b$, and the number of observations is $N$. Table \ref{table2} preliminarily lists the information of RSS input for different algorithms. More details of experimental settings are presented as follows:

\textbf{RadioGAT}: 
The blocks in $\mathbf{B_1}$ are used as a training set to train the proposed graph attention network, and then RadioGAT is used to predict the RSS of the target frequency. First, for blocks in $\mathbf{B}$, the graph of $f_{obv}$ and $f_{target}$ are constructed based on the introduced model-based spatial-spectral encoding methods. Then the training procedure is executed as section V. After $200$ epochs of training at a learning rate of $0.001$, the trained RadioGAT can perform prediction from $p_{b,Nobv}^{(obv)}$ to $p_{b,N}^{(target)}$. In the inference procedure, the sparse observation $p_{b,Nobv}^{(k)}$, node positions, $f_{obv}$ and $f_{target}$ are input into RadioGAT across different channels as $X_b$ for block $b$ in $\mathbf{B_2}$. $\mathcal{E}_{b}^{obv}$ and $\mathcal{E}_{b}^{target}$ are input into GAT layer 1 and the remaining GAT layers as the graph adjacency matrix, respectively. Finally, the whole network outputs $\hat{p}_{b,N}^{(target)}$ as the prediction of $p_{b,N}^{(target)}, \forall b \in \mathbf{B_2}$. To facilitate comparison with other algorithms, we set $f_{obv}$ and $f_{target}$ to 1750MHz and 4750MHz/5750MHz.

\begin{figure*}[t]
\centering

\subfloat[1: GroundTruth]{\includegraphics[scale=0.235]{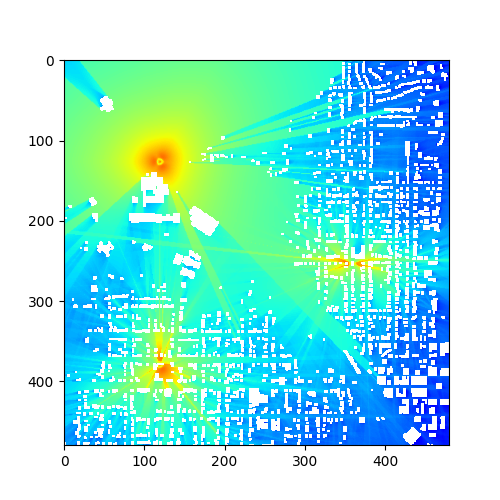}}
\subfloat[2: GroundTruth]{\includegraphics[scale=0.235]{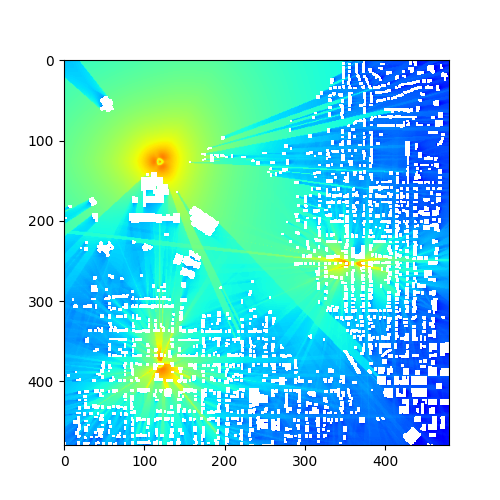}}
\subfloat[3: GroundTruth]{\includegraphics[scale=0.235]{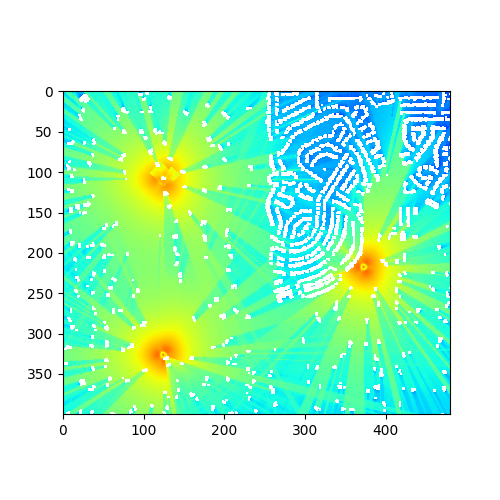}}
\subfloat[1: RadioGAT\newline \centering {RMSE: 2.4062 dB}]{\includegraphics[scale=0.235]{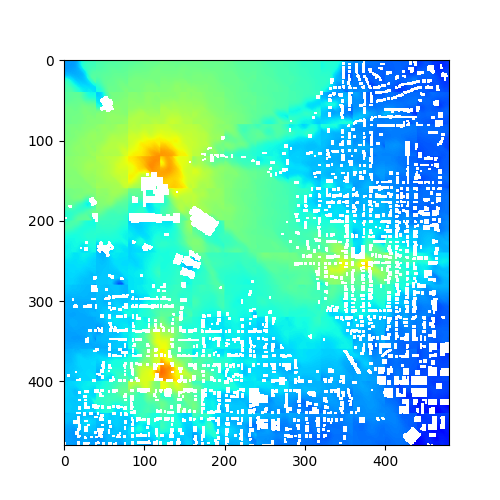}}
\subfloat[2: RadioGAT\newline \centering {RMSE: 2.4659 dB}]{\includegraphics[scale=0.235]{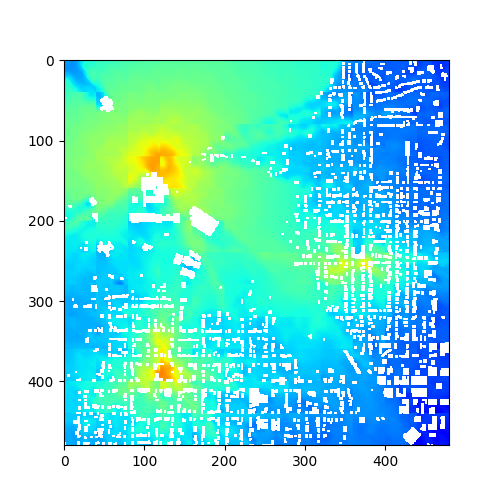}}
\subfloat[3: RadioGAT\newline\centering {RMSE: 2.7396 dB}]{\includegraphics[scale=0.235]{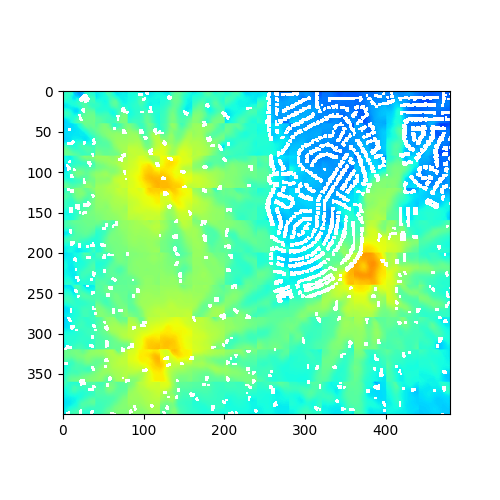}}
\\
\subfloat[1: GNN\newline \centering {RMSE: 3.4545 dB}]{\includegraphics[scale=0.235]{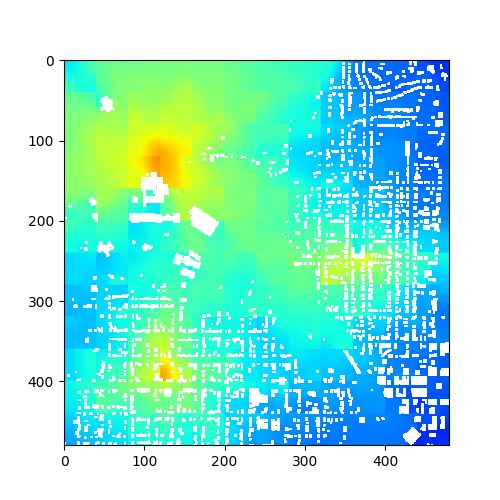}}
\subfloat[2: GNN\newline \centering {RMSE: 2.7801 dB}]{\includegraphics[scale=0.235]{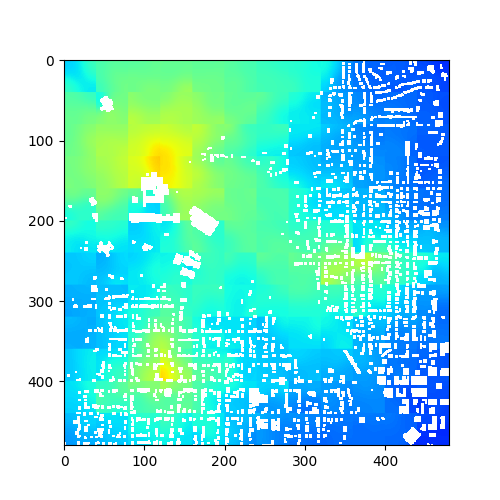}}
\subfloat[3: GNN\newline \centering {RMSE: 3.7347 dB}]{\includegraphics[scale=0.235]{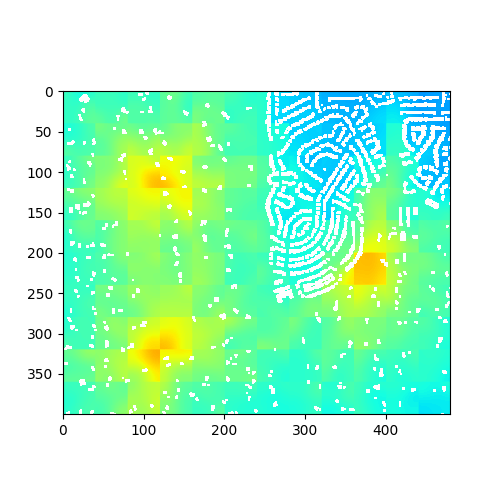}}
\subfloat[1: Autoencoder\newline\centering {RMSE: 7.4560 dB}]{\includegraphics[scale=0.235]{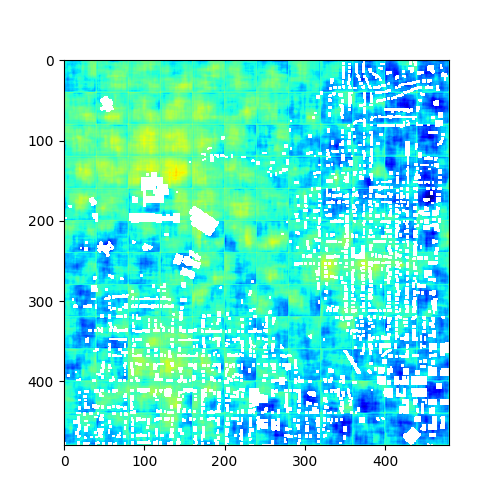}}
\subfloat[2: Autoencoder\newline \centering {RMSE: 7.4814 dB}]{\includegraphics[scale=0.235]{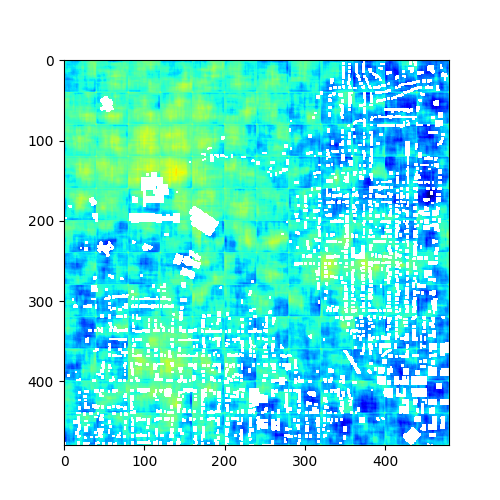}}
\subfloat[3: Autoencoder\newline \centering {RMSE: 7.9314 dB}]{\includegraphics[scale=0.235]{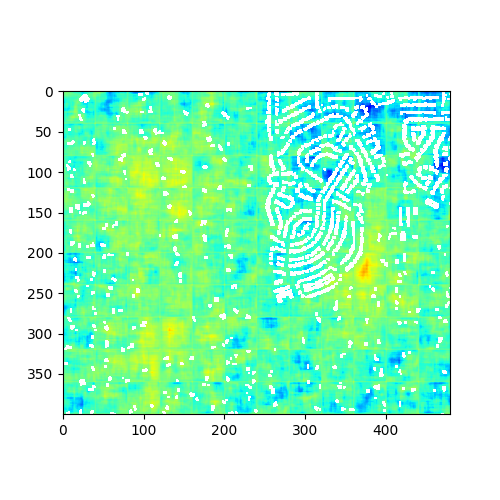}}
\\
\subfloat[1: CGAN\newline \centering {RMSE: 11.2629 dB}]{\includegraphics[scale=0.235]{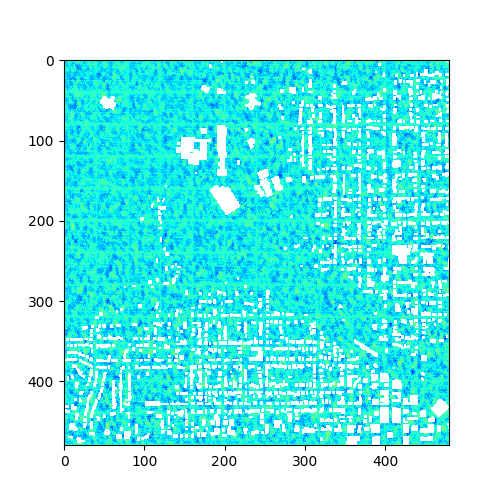}}
\subfloat[2: CGAN\newline \centering {RMSE: 13.7752 dB}]{\includegraphics[scale=0.235]{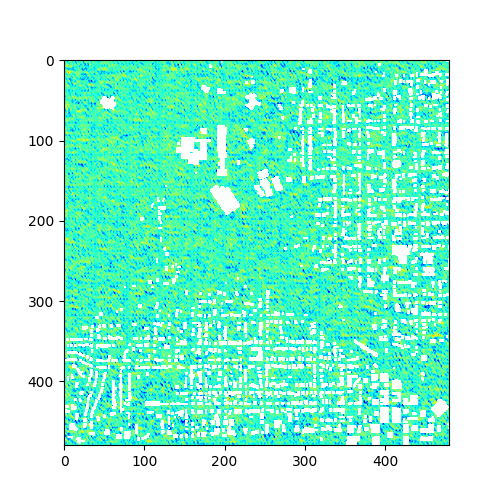}}
\subfloat[3: CGAN\newline \centering {RMSE: 13.7200 dB}]{\includegraphics[scale=0.235]{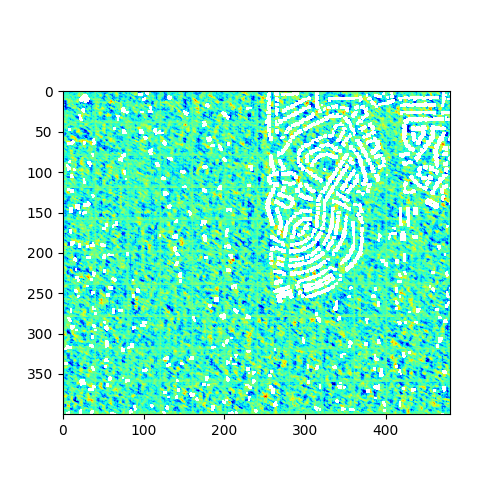}}
\subfloat[1: SF-Kriging\newline \centering {RMSE: 3.4834 dB}]{\includegraphics[scale=0.235]{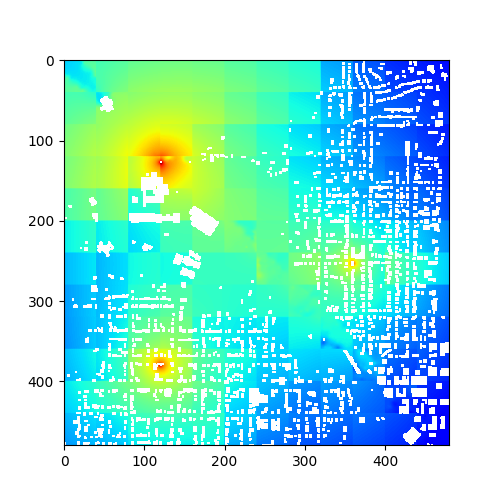}}
\subfloat[2: SF-Kriging\newline \centering {RMSE: 3.5862 dB}]{\includegraphics[scale=0.235]{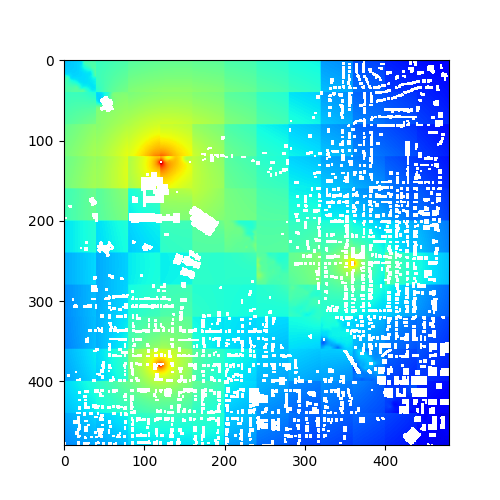}}
\subfloat[3: SF-Kriging\newline \centering {RMSE: 3.9736 dB}]{\includegraphics[scale=0.235]{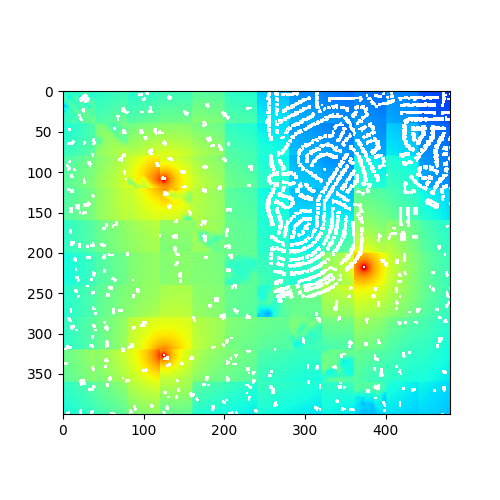}}
\\
\subfloat[1: 3D-IDW \newline \centering {RMSE: 4.5184 dB}]{\includegraphics[scale=0.235]{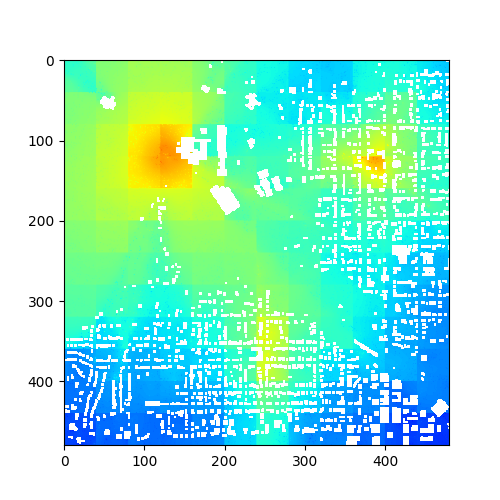}}
\subfloat[2: 3D-IDW\newline \centering {RMSE: 6.0974 dB}]{\includegraphics[scale=0.235]{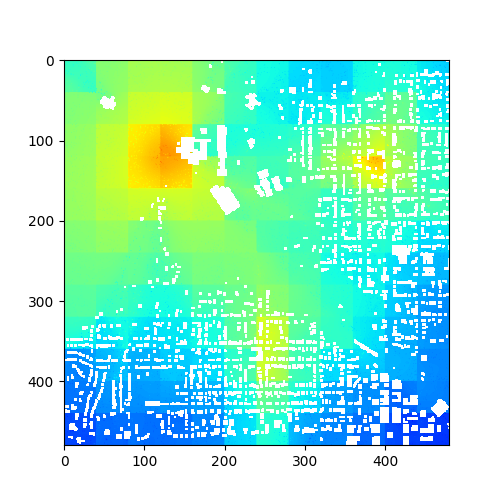}}
\subfloat[3: 3D-IDW\newline \centering {RMSE: 5.2633 dB}]{\includegraphics[scale=0.235]{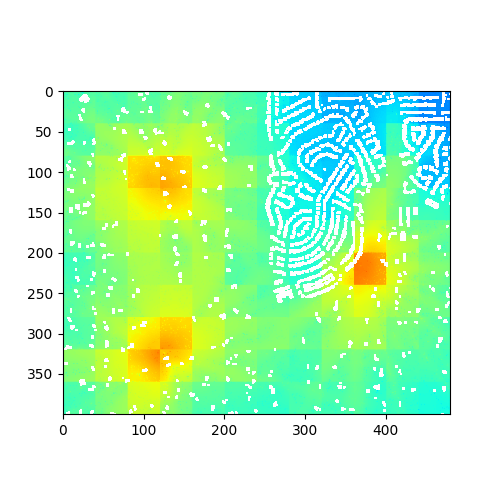}}
\subfloat[1: TensorCompletion\newline \centering {RMSE: 11.0177 dB}]{\includegraphics[scale=0.235]{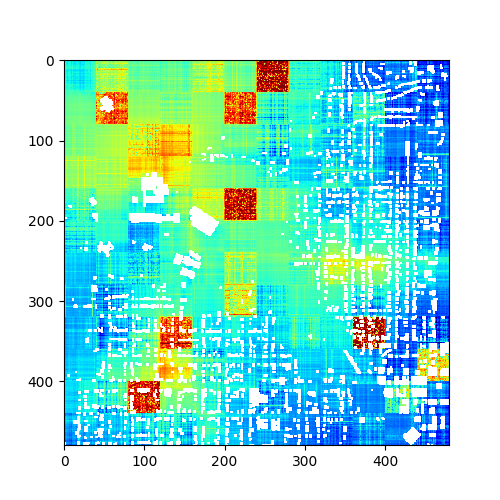}}
\subfloat[2: TensorCompletion\newline \centering {RMSE: 11.3420 dB}]{\includegraphics[scale=0.235]{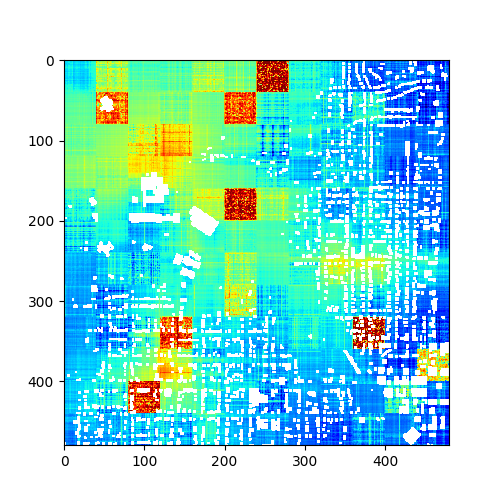}}
\subfloat[3: TensorCompletion\newline 
 \centering {RMSE: 9.7585 dB}]{\includegraphics[scale=0.235]{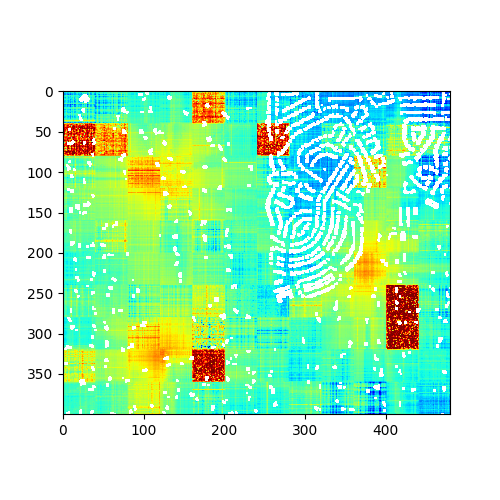}}
\\
{\includegraphics[scale=0.81]{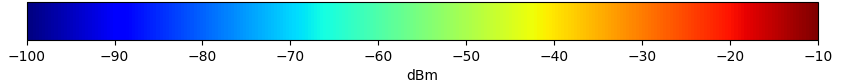}}
\caption{The prediction radiomap visualization results for different algorithms under different conditions {(1: area 6 at 4750MHz, 2: area 6 at 5750MHz, 3: area 10 at 4750MHz).}}
\label{fig8}
\end{figure*}
\begin{figure*}[t]
\centering
\captionsetup{labelfont={blue},textfont={blue}}
\begin{minipage}{0.48\linewidth}
\centering
\includegraphics[width=0.8\linewidth]{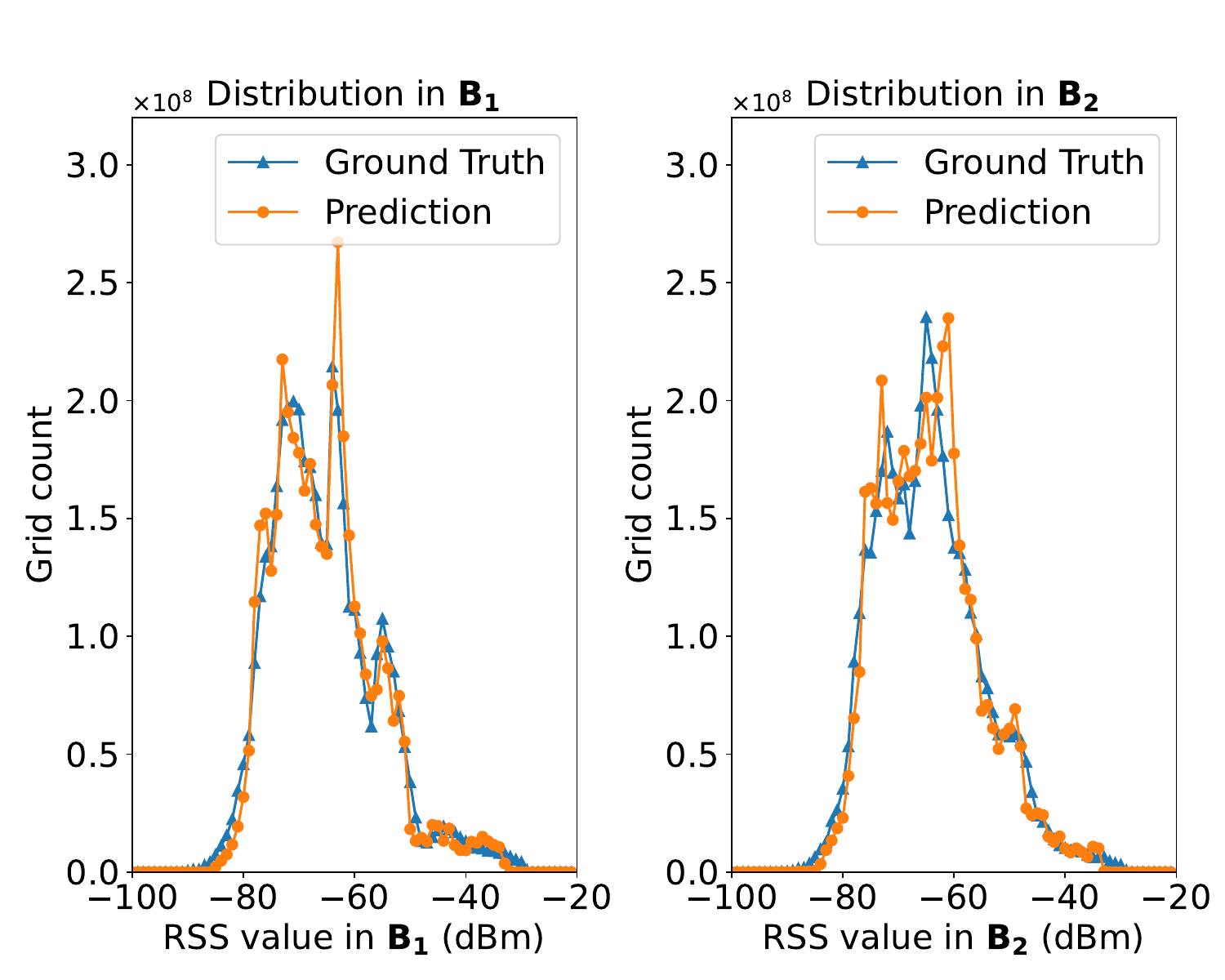}
\caption{A histogram example in reconstructed radiomaps, where the left picture represents the distribution fitting result of the training set, and the right picture represents the distribution fitting result of the test set.}
\label{fig9}
\end{minipage}
\hspace{2mm}
\begin{minipage}{0.48\linewidth}
\centering
\includegraphics[width=0.8\linewidth]{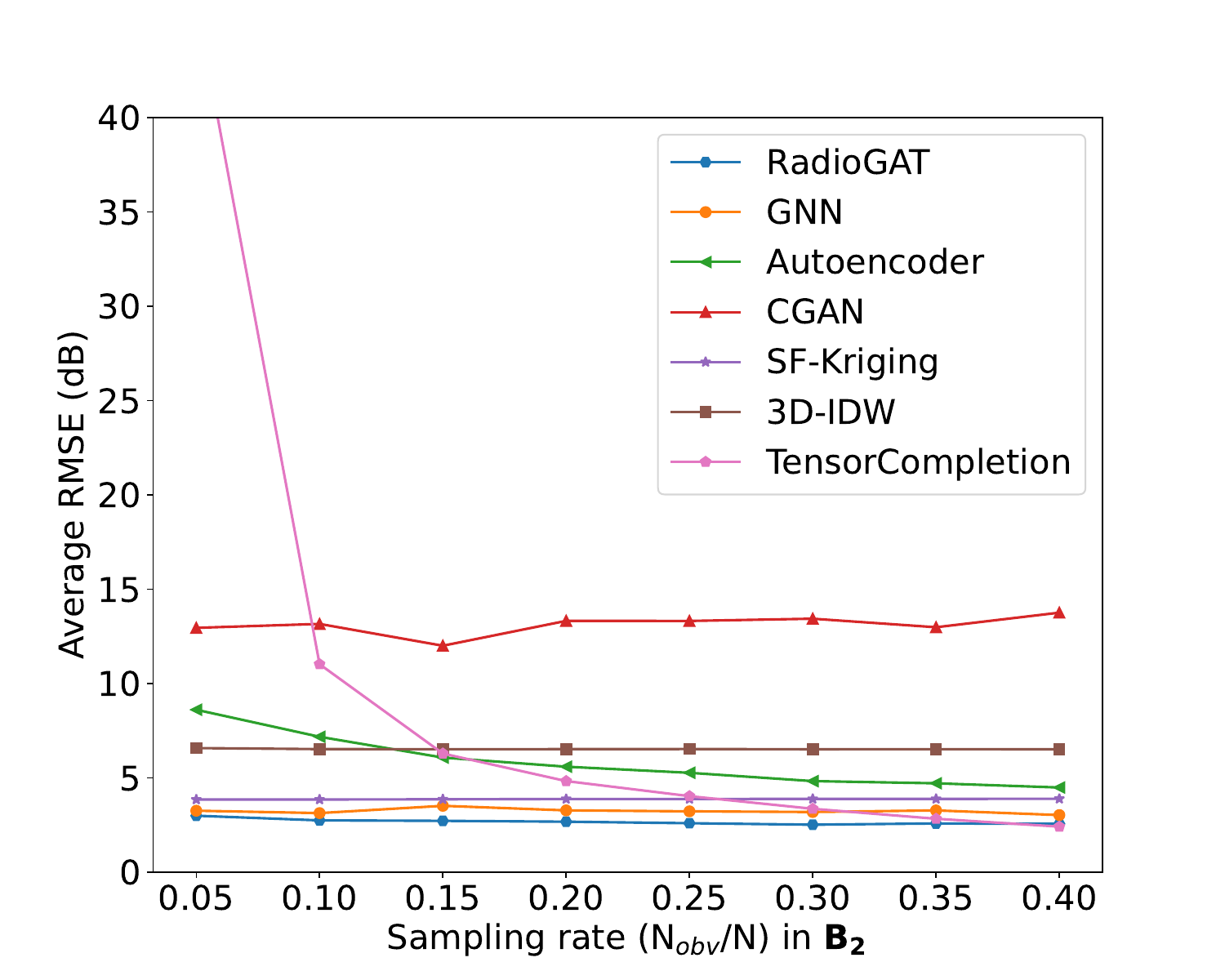}
\caption{The average RMSE for different algorithms in all areas under different $N_{obv}/N$.}
\label{fig10}
\end{minipage}
\end{figure*}

\textbf{GNN}: 
Set the blocks in $\mathbf{B_1}$ as a training set. The inputs are the sparse RSS observation $p_{b,Nobv}^{(1750)}$ and the adjacency matrix containing distance information to the network. Two GNNs with prediction outputs of $p_{b,N}^{(4750)}$ and $p_{b,N}^{(5750)}$ are trained separately, and they performs separately prediction of $f_{target}$ in $\mathbf{B_2}$.

\textbf{Autoencoder}:
The blocks in $\mathbf{B_1}$ are used as a training set to train the autoencoder network. The network input is a three-dimensional tensor containing sparse observation ($p_{b,Nobv}^{(1750)}$, $p_{b,Nobv}^{(2750)}$ and $p_{b,Nobv}^{(3750)}$, $\forall b \in \mathbf{B_1}$) while the output is a three-dimensional tensor containing complete observation ($p_{b,N}^{(1750)}$, $p_{b,N}^{(2750)}$, $p_{b,N}^{(3750)}$ and $p_{b,N}^{(target)}$,$\forall b \in \mathbf{B_1}$). 
After $200$ epochs of training at a learning rate of $0.001$, the trained autoencoder is used to predict the RSS at $f_{target}$ in $\mathbf{B_2}$.

\textbf{cGAN}:
The blocks in $\mathbf{B_1}$ are used as a training set to train the cGAN. The generator in cGAN predicts complete observations under Gaussian white noise and sparse observation input ($p_{b,Nobv}^{(1750)}$, $p_{b,Nobv}^{(2750)}$ and $p_{b,Nobv}^{(3750)}$, $\forall b \in \mathbf{B_1}$), while the discriminator takes sparse observations and predicted values as input to determine whether the input is true or false. After $200$ epochs of training at a learning rate of $0.001$, the generator is used to predict completed observations $p_{b,N}^{(target)}$ in $\mathbf{B_2}$.

\textbf{SF-Kriging}: For block b in $\mathbf{B_1}$, $p_{b,N}^{(1750)}$ and $p_{b,N}^{(target)}$ are used to estimate large-scale fading signal parameters under the least squares method. For blocks in $\mathbf{B_2}$, $p_{b,Nobv}^{(1750)}$ is used to estimate the shadow fading. Finally, the estimated large-scale fading parameters and shadow fading are used to predict $p_{b,n}^{(target)}$ in $\mathbf{B_2}$.

\textbf{3D-IDW}: For block b in $\mathbf{B_2}$, $p_{b,Nobv}^{(1750)}$, $p_{b,Nobv}^{(2750)}$ and $p_{b,Nobv}^{(3750)}$, $\forall b \in \mathbf{B_2}$ are observable. 3D-IDW exploits the spatial and frequency distance between grids to predict the unobserved radio signal strength. The predicted value under 3D-IDW is used as $p_{b,n}^{(target)}$ in $\mathbf{B_2}$.

\textbf{TensorCompletion}: We splice $p_{b,Nobv}^{(1750)}$, $p_{b,Nobv}^{(2750)}$, $p_{b,Nobv}^{(3750)}$ and $p_{b,Nobv}^{(target)},\forall b \in \mathbf{B_2}$ as a three-dimensional tensor. The HaLRTC algorithm in \cite{b41} is used to complete the three-dimensional tensor completion task. The number of iterations and $\rho$ for HaLRTC are respectively set as $1500$ and $10^{-6}$.


\subsection{Performance Evaluation}

To evaluate the performance of different algorithms, all areas shown in TABLE \ref{table1} are used. The size of the block is set to $ 200\times 200 \mathrm{m^2}$. Each area is randomly divided into $\mathbf{B_1}$ ($N_{B1}$ blocks in $\mathbf{B_1}$) and $\mathbf{B_2}$ ($N_{B2}$ blocks in $\mathbf{B_2}$) under $N_{B1}/N_{B}=50\%$. Then the inference RMSE in $\mathbf{B_2}$ is utilized to quantify the performance at each area. After denormalizing ${y}_{b,n}$ and $\hat{y}_{b,n}$ back to RSS, ${y}_{b,n}$ and $\hat{y}_{b,n}$ are retained to denote the result post-denormalization. The RMSE in $b, \forall b \in \mathbf{B_2}$ can be calculated as
\begin{equation}
    \mathrm{RMSE}_{b}=\sqrt{\frac{1}{N} \sum_{n=1}^{N} \left(y_{b,n}-\hat{y}_{b,n}\right)^{2}}.
\end{equation}
The whole RMSE in each area can be calculated as
\begin{equation}
    \mathrm{RMSE}=\sqrt{ \frac{1}{N_{B2}} \sum_{b \in \mathbf{B_2}} \mathrm{RMSE}_{b}^2}. 
\end{equation}
Furthermore, the average RMSE in all areas stands for the algorithm performance. To show the superiority of RadioGAT, we conduct supervised learning (100\% masked) experiments, evaluating the performance of different algorithms in \textit{1)} - \textit{5)}. A semi-supervised learning experiment (5\% masked) is conducted to demonstrate the generalizability and deployment feasibility of RadioGAT as illustrated in \textit{6)}.

\subsubsection{Evaluation of Different Correlation Encoding Methods}\label{subsub1}
Before the comparison to SOTA approaches, we first evaluate the performance of different correlation encoding methods for RadioGAT, as
 shown in TABLE \ref{table3}. Specifically, we evaluate the radiomap prediction RMSE under different correlation encoding methods in each area. { A validation dataset can be divided from the training set for appropriate selection of hyperparameters. More specifically,} in our experiments, $d_{th}$, $\delta$, and $\beta$ are set to $15$m, $1 \times 3$ (subject to $M=3$) and $10$. The average RMSE in all correlation encoding methods is below 3.8 dB, and the highest RMSE in all areas does not exceed 6 dB. This shows that the proposed RadioGAT scheme can maintain high prediction accuracy under various correlation encoding methods. In addition, our proposed model-based spatial-spectral encoding method achieves the lowest average RMSE in ten areas, and the predicted RMSE was the best in most areas. 
 {Generally, RadioGAT achieves good performances under different edge encoding methods, which demonstrates its generalization. However, the design of edge encoding methods still impacts the performance. From the results, optimal performance is achieved using our proposed model-based method by taking into account both spatial and frequency correlations of grid nodes.}

\subsubsection{RMSE Performance}


To further demonstrate the superiority of the proposed scheme. We compare model-based spatial-correlation correlation encoding RadioGAT (hereafter abbreviated as RadioGAT) with other benchmarks. Table \ref{table4} shows the RMSE (dB) results of the seven algorithms constructed in ten areas. According to the average RMSE comparison results in ten areas, the prediction of 4750MHz and 5750MHz under the proposed RadioGAT is 2.8384dB and 2.9931dB respectively, achieving optimal RMSE performance compared to other algorithms.

\subsubsection{Visualization Results}
Beyond numerical results, we also provide the visualization results as shown in Fig. \ref{fig8}.
Here we consider three conditions: 1) the prediction of area 6 at 4750MHz; 2) the prediction of area 6 at 5750MHz and 3) the prediction of area 10 at 4750MHz. For TensorCompletion, $N_{obv}/N=10\%$ while for other algorithms, $N_{obv}/N=5\%$. The visualization results are spliced by the prediction results of the target frequency radiomap of all blocks by different algorithms. {With the integration of model-based and data-driven methods, our proposed RadioGAT takes advantage of both data statistics and physical radio propagation principles, leading to a more accurate radiomap reconstruction and smoother patterns, which is consistent with our numerical results.}

As shown in Fig. \ref{fig9}, we also plot the histogram of the prediction radiomap of area 10 at 5750MHz for all blocks. The result shows that the proposed algorithm captures the distribution of the radiomap in the area, which further validates the efficiency of RadioGAT.

\subsubsection{Impact of Grids Sampling Rate $N_{obv}/N$} 

We show the performance under different sampling rates at each block in Fig. \ref{fig10}. We test the average RMSE of 5750MHz radiomap prediction in 10 areas when $N_{obv}/N$ is equal to $[5\%,10\%,15\%,20\%,25\%,30\%,35\%,40\%]$. In addition, to ensure that the prior information is under the same conditions, sampling grids of the low sampling rate are sampled from the subset of $N_{obv}/N=40\%$. The performance of the TensorCompletion and autoencoder improves with increasing $N_{obv}/N$. This shows that, as $N_{obv}/N$ increases, the neural networks gradually fit a specific distribution while the performance may decrease if the distribution changes \cite{b39}. The performance of SF-Kriging and 3D-IDW keeps similar with different $N_{obv}/N$. We additionally test the performance when $N_{obv}/N$ is lower than $5\%$ and found that SF-Kriging and 3D-IDW reach convergence at about $N_{obv}/N=1\%$. Moreover, the performance of cGAN remains low accuracy due to insufficient training samples (the number of training blocks is only about $50$). Though the RMSE of GNN remains low, it cannot utilize sampling information from multiple frequency bands simultaneously. Overall, our proposed RadioGAT achieves the best performance when $N_{obv}/N$ is lower than $40\%$. It is worth noting that when $N_{obv}/N=40\%$, TensorCompletion performs the best, which is usually impractical in realistic scenarios. {Guided by the knowledge of the physical radio propagation model, RadioGAT only uses a small amount of data to train the network and achieve optimal performance.}

\begin{figure}[t]
\centering
\captionsetup{labelfont={blue},textfont={blue}}
\subfloat[Minimum RMSE]{\includegraphics[scale=0.3]{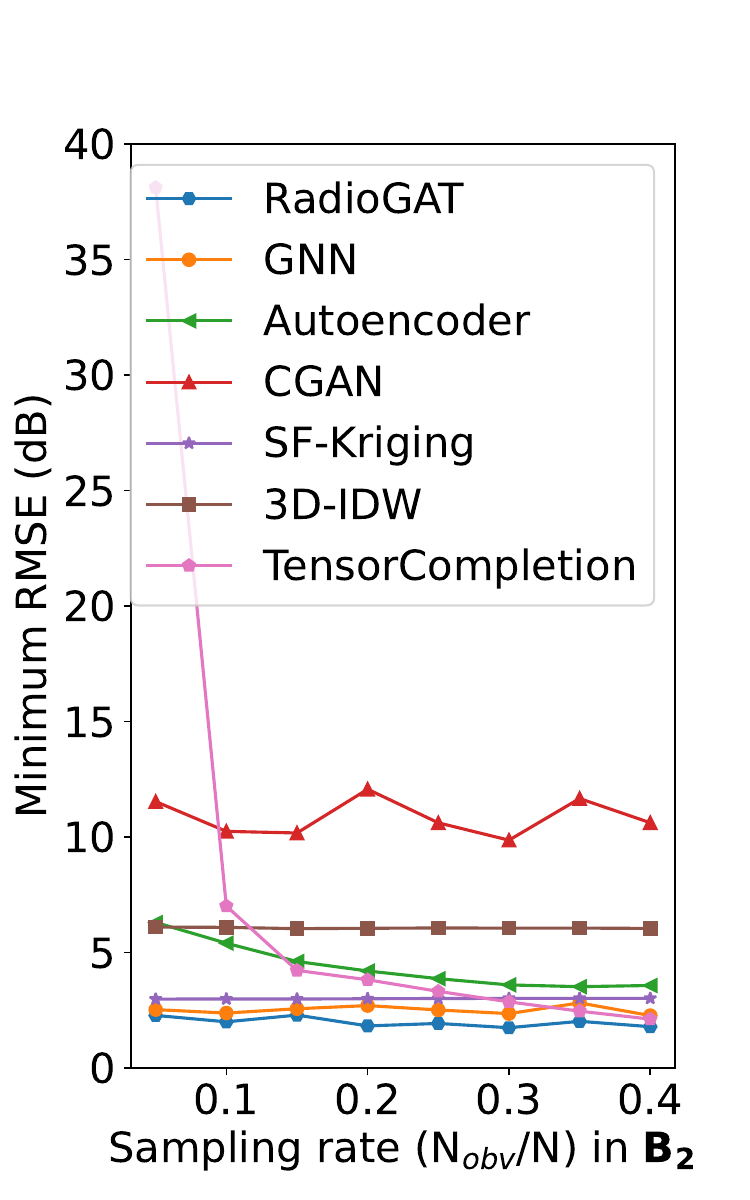}}
\subfloat[Maximum RMSE]{\includegraphics[scale=0.3]{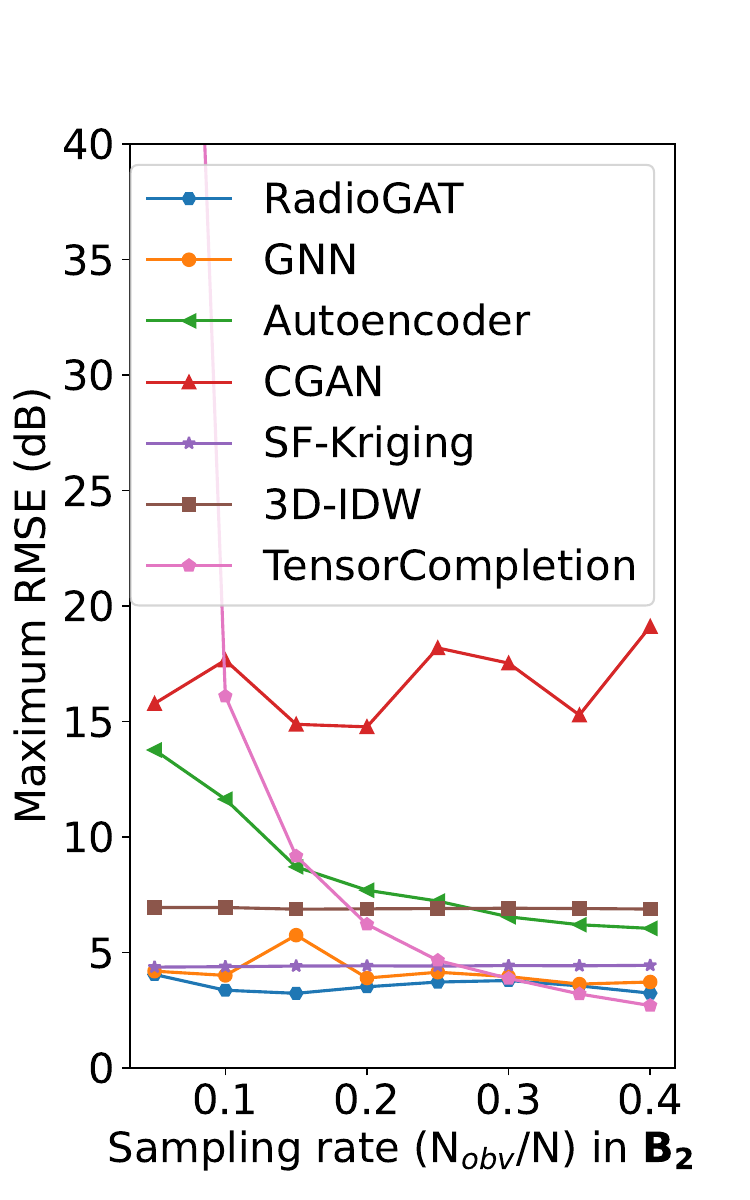}}
\caption{The minimum and maximum RMSE for different algorithms in all areas under different $N_{obv}/N$.}
\label{fig11}
\end{figure}
\begin{figure}[t]
\captionsetup{labelfont={blue},textfont={blue}}
\centering

{\includegraphics[scale=0.31]{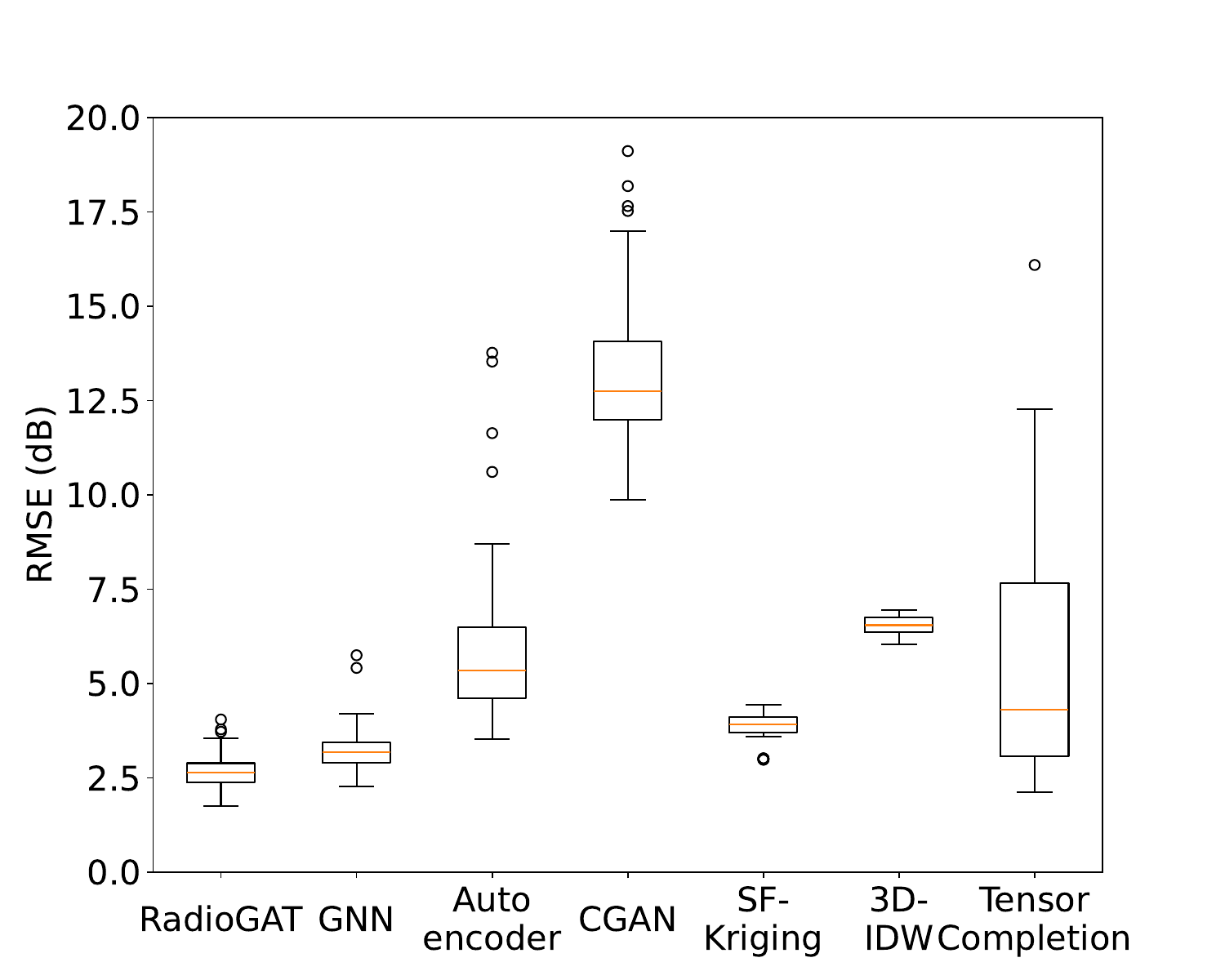}}
\caption{The boxplot of RMSE performance in different areas, frequencies, and sampling rates $N_{obv}/N$ for different algorithms.}
\label{figBoxplot}
\end{figure}
\begin{table}[t]
\caption{\centering{RMSE with Semi-supervised Learning}}
\centering
\label{tablesemi}
\begin{tabular}{|cc|c|cc|c|}
\hline
\multicolumn{1}{|c|}{Area}               & f/MHz & RMSE (dB)   & \multicolumn{1}{c|}{Area}                & f/MHz & RMSE (dB)   \\ \hline
\multicolumn{1}{|c|}{\multirow{2}{*}{1}} & 4750  & 3.3054 & \multicolumn{1}{c|}{\multirow{2}{*}{6}}  & 4750  & 2.6742 \\ \cline{2-3} \cline{5-6} 
\multicolumn{1}{|c|}{}                   & 5750  & 3.5979 & \multicolumn{1}{c|}{}                    & 5750  & 2.7719 \\ \hline
\multicolumn{1}{|c|}{\multirow{2}{*}{2}} & 4750  & 3.2362 & \multicolumn{1}{c|}{\multirow{2}{*}{7}}  & 4750  & 5.5707 \\ \cline{2-3} \cline{5-6} 
\multicolumn{1}{|c|}{}                   & 5750  & 3.8801 & \multicolumn{1}{c|}{}                    & 5750  & 3.8057 \\ \hline
\multicolumn{1}{|c|}{\multirow{2}{*}{3}} & 4750  & 2.836  & \multicolumn{1}{c|}{\multirow{2}{*}{8}}  & 4750  & 3.9791 \\ \cline{2-3} \cline{5-6} 
\multicolumn{1}{|c|}{}                   & 5750  & 2.6334 & \multicolumn{1}{c|}{}                    & 5750  & 4.5976 \\ \hline
\multicolumn{1}{|c|}{\multirow{2}{*}{4}} & 4750  & 3.3476 & \multicolumn{1}{c|}{\multirow{2}{*}{9}}  & 4750  & 3.5491 \\ \cline{2-3} \cline{5-6} 
\multicolumn{1}{|c|}{}                   & 5750  & 2.8966 & \multicolumn{1}{c|}{}                    & 5750  & 6.2819 \\ \hline
\multicolumn{1}{|c|}{\multirow{2}{*}{5}} & 4750  & 2.7146 & \multicolumn{1}{c|}{\multirow{2}{*}{10}} & 4750  & 3.1317 \\ \cline{2-3} \cline{5-6} 
\multicolumn{1}{|c|}{}                   & 5750  & 6.8445 & \multicolumn{1}{c|}{}                    & 5750  & 3.274  \\ \hline
\multicolumn{2}{|c|}{Average (4750)}             & \textbf{3.4345} & \multicolumn{2}{c|}{Average (5750)}              & \textbf{4.0584} \\ \hline
\end{tabular}
\end{table}
\begin{figure}[t]
\centering
\subfloat[\centering {RMSE: 2.6742 dB}]{\includegraphics[scale=0.23]{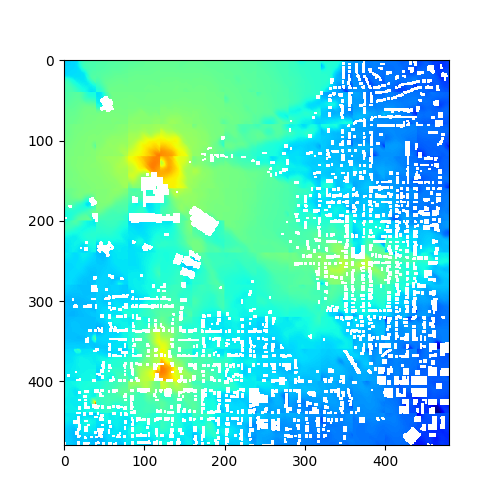}}
\subfloat[\centering {RMSE: 2.7719 dB}]{\includegraphics[scale=0.23]{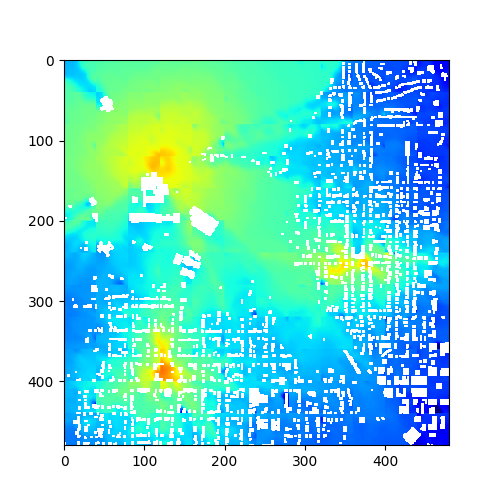}}
\subfloat[\centering {RMSE: 3.1317 dB}]{\includegraphics[scale=0.23]{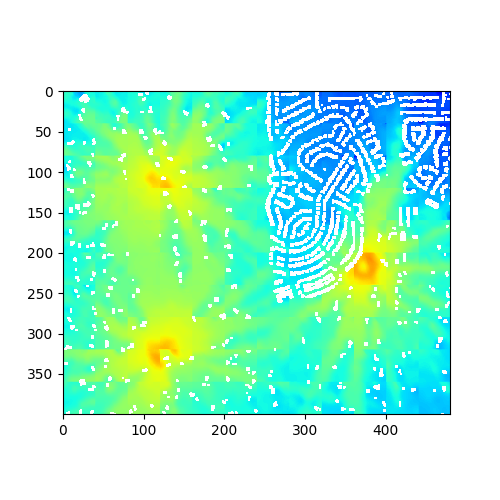}}
\caption{The visualization result {of RadioGAT} under semi-supervised learning setting for three conditions: {(a) for area 6 at 4750MHz, (b) for area 6 at 5750MHz, and (c) for area 10 at 4750MHz.}}
\label{figsemi}
\end{figure}
\subsubsection{Algorithm Robustness}
{To evaluate the robustness of different algorithms, the minimum and maximum RMSE for all areas at 5750MHz under different sampling rates are shown in Fig. \ref{fig11}. RadioGAT achieves the best performance when sampling rate $N_{obv}/N$ is less than $35\%$. It is noted that the maximum RMSE of TensorCompletion reaches best when sampling rate $N_{obv}/N$ is more than $35\%$. However, such a high sampling rate is usually inaccessible in practical scenarios. To better illustrate the robustness of RadioGAT, we further present the performance of different algorithms with different areas, frequencies, and sampling rates in the boxplot, as shown in Fig. \ref{figBoxplot}. The superior results of RadioGAT further demonstrate the robustness brought by the model-based design in the proposed algorithm.}

\subsubsection{Semi-Supervised Learning with RadioGAT}
Our evaluations extend to semi-supervised learning scenarios where RadioGAT demonstrates robust performance with limited label availability. Specifically, when only 5\% of the observations at frequency $f_{target}$ for the block set $\mathbf{B_1}$ are available (5\% masked), with a learning rate of 0.005 and 200 training epochs, RadioGAT sustains an average RMSE below 4.1dB, as delineated in Table \ref{tablesemi} and illustrated in Fig. \ref{figsemi}. The results show RadioGAT's capability to yield reasonably accurate estimates even with sparse ground truth data. {This further reveals the effectiveness of graph structure and masked training techniques. In addition, the block in this paper is set as a square to facilitate comparison with CNN-based algorithms. Due to the use of graph structure, the block does not need to be fixed as a square when applying RadioGAT in an actual scenario, which could further enhance the flexibility of RadioGAT.}


\section{Conclusion} \label{sec: conclusion}
In this work, we introduce a novel framework, namely RadioGAT, for multi-band radiomap reconstruction (MB-RMR) via Graph Attention Network (GAT), which consists of two phases: 1) a model-based method for encoding spatial-spectral correlations within a graph structure, and 2) a data-driven approach using GAT for cross-band radiomap generalization. {The integration of model-based and data-driven methods brings in high-precision radiomap reconstruction capabilities in RadioGAT, even with a small amount of observed samples.}
Experimental results demonstrate RadioGAT's efficacy and accuracy in supervised learning contexts, as well as its adaptability in semi-supervised, data-sparse environments.

{With the deployment of IoTs and mobile devices\cite{zhu2023pushing,wen2023task}, efficient radiomap estimation becomes increasingly important for spectrum management and network optimization. One critical challenge is to develop fast radiomap estimation to enable real-time spectrum analysis. Despite the superior performance of the proposed RadioGAT in accuracy, more efforts shall be contributed to addressing the trade-off of computational complexity and performance. Particularly, the inter-block correlations shall be investigated to increase the MB-RMR efficiency. Another promising future direction is to predict new spectrum coverage from current bands, such as estimating 5G from 4G data and incorporating it into spectrum management for 6G technology. Other potential research topics also include the exploration of integrating generative AI and advanced large language models to improve multi-band radiomap estimation.}

\ifCLASSOPTIONcaptionsoff
  \newpage
\fi





\bibliographystyle{IEEEtran}
\bibliography{IEEEabrv,Bibliography}

\vfill


\end{document}